\documentclass[letterpaper,11pt]{article}
\pdfoutput=1

\usepackage{jheppub}
\usepackage{comment}

\usepackage{hyperref} 
\usepackage{epsfig}
\usepackage[normalem]{ulem}
\usepackage{bm}
\usepackage{bbm}
\usepackage{slashed}
\usepackage{xspace}
\usepackage{subfigure}
\usepackage{multirow}
\usepackage{amsmath,latexsym}
\usepackage{pstricks}
\usepackage{color} 

\newcommand{\setmainskip}{\setlength\baselineskip{18pt}}

\usepackage{marginnote}

\newcommand{\eq}[1]{Eq.~\eqref{eq:#1}}
\newcommand{\eqs}[2]{Eqs.~\eqref{eq:#1} and \eqref{eq:#2}}

\newcommand{\fig}[1]{Fig.~\ref{fig:#1}}

\newcommand{\nn}{\nonumber}

\newcommand{\beq}{\begin{equation}}
\newcommand{\eeq}{\end{equation}}
\newcommand{\bea}{\begin{eqnarray}}
\newcommand{\eea}{\end{eqnarray}}

\def\psip#1{\psi_{\mathbf{#1}}}

\def\psipd#1{\psi^\dagger_{\mathbf{#1}}}

\def\bsigma{\mbox{\boldmath $\sigma$}}

\newcommand{\bmp}{\mathbf p}

\newcommand{\ppp}{\mbox{$({\mathbf p'}-{\mathbf p})^2$}}
\newcommand{\spp}{\mbox{$i\bsigma\cdot (\mathbf{ p'\times p})$}}

\renewcommand{\tabcolsep}{1ex}
						
\allowdisplaybreaks[1]

\begin{document}

\preprint{
\hbox{MIT--CTP 4892}
}
							
\title{\boldmath Manifestly Soft Gauge Invariant Formulation of vNRQCD}

\author{Ira Z. Rothstein$^1$, Prashant Shrivastava$^1$ and Iain W. Stewart$^2$}
							
\affiliation{$^1$Department of Physics, Carnegie Mellon University, Pittsburgh, PA 15213, USA}
\affiliation{$^2$Center for Theoretical Physics, Massachusetts Institute of Technology, Cambridge, MA 02139, USA}
							
\emailAdd{izr@andrew.cmu.edu}
\emailAdd{prashans@andrew.cmu.edu}
\emailAdd{iains@mit.edu}
							
\abstract{
								
\vspace{0.2cm}
\setlength\baselineskip{15pt}
							
Homogeneous power counting in Non-Relativistic QCD (NRQCD) implies the simultaneous existence of both soft and ultrasoft gluons.  In the velocity renormalization group (vNRQCD) formalism we show that operators involving soft fields interacting with heavy potential quarks can be put in a manifestly gauge invariant form by utilizing gluon and quark building blocks which contain Wilson lines, and are analogous to those used in the soft collinear effective theory.   This leads to several simplifications, in particular significantly reducing the size of the operator basis, which stream-lines  matching and anomalous dimension calculations at subleading order in the velocity expansion. Also, soft ghosts no longer couple via potential like interactions to the heavy quark fields, and hence do not appear until two loops. Furthermore the  the color structures that appear at each order in $\alpha_s$  in the static potential are only those which should arise according to non-Abelian exponentiation. We also discuss how zero-bin subtractions clarify the role of $i\varepsilon$ terms in the heavy quark propagator poles carrying soft momenta. Even though the choice of the direction of the soft Wilson lines in our building blocks does not matter, there is still a limited set of consistent choices, and the predicted form of the eikonal poles does influence results at two-loops and beyond.

}
							

\maketitle
\setmainskip

\newpage

\section{Introduction}

Non Relativistic QCD (NRQCD) is an effective field theory (EFT), proposed in its original form in  \cite{Caswell:1985ui,Bodwin:1994jh},  which has been successfully used to describe the dynamics of non-relativistic heavy quark-antiquark bound states. This EFT includes  three scales which are the heavy quark mass $m$, momentum $mv$ and energy $mv^2$. What complicates NRQCD compared to more canonical effective field theories is that two infrared scales, i.e energy and momentum, are not independent. They are instead related by the free dispersion relation, a fact encoded by carrying out the power expansion in the heavy quark relative velocity, and correspondingly the logarithms of $p$ and $E$ are in general not independent.  A systematic way to deal with these coupled scales was proposed in~\cite{Luke:1999kz}, see~\cite{Rothstein:2003mp, Hoang:2002ae} for reviews and~\cite{Luke:1996hj,Manohar:1997qy,Grinstein:1997gv,Pineda:1997bj,Beneke:1997zp,Griesshammer:1997wz} for earlier foundational work. In this formulation, usually called ``vNRQCD", one carries out the matching in a single stage, and simultaneously sum logs of both the energy and momentum by using what is called the ``velocity renormalization group", as has been carried out in Refs.~\cite{Manohar:1999xd,Manohar:2000kr,Manohar:2000rz,Manohar:2000mx,Hoang:2001rr,Hoang:2002yy,Hoang:2003ns,Hoang:2006ht,Hoang:2011gy}.  The framework of pNRQCD provides an alternative  formulation of the EFT where the soft and ultrasoft scales are treated independently--for more of the setup, see Refs.~\cite{Pineda:1997bj,Pineda:1998kn,Brambilla:1999xf,Brambilla:1999qa,Pineda:2000gz,Pineda:2001et,Pineda:2001ra,Brambilla:2004jw,Pineda:2011aw}.

While the present formulation of vNRQCD in \cite{Luke:1999kz} is quite useful for summing logs, and yields correct results,  the operators involving the interaction of soft gauge fields and heavy quarks are not manifestly soft gauge invariant. This shortcoming of the formalism makes computing the soft contribution to the running  of the subleading potentials  cumbersome, as can be seen in Refs.~\cite{Manohar:1999xd,Manohar:2000kr}. Although the final results for these potentials are  gauge invariant, this lack of  manifest symmetry proliferates the number of integrals that need to be calculated. It should be expected that if we instead formulate the theory in a manifestly gauge invariant way (in both the soft and ultrasoft sectors), then simplifications will follow.

In this paper, following ideas put forward in \cite{Rothstein:2016bsq} in the context of forward scattering of energetic particles in the soft collinear effective theory~\cite{Bauer:2000ew,Bauer:2000yr,Bauer:2001ct,Bauer:2001yt}, we formulate NRQCD so as to maintain
soft gauge invariance at all stages of the calculation. This is accomplished by 
including soft Wilson lines in operators that connect soft and potential fields in the action through gauge invariant soft gluon and quark building block fields. We demonstrate the utility of this formalism by showing
how it simplifies the calculation of the anomalous dimensions of the heavy quark potential to order $O(v^2)$.  We also show how our formalism, in conjunction with the vNRQCD zero-bin subtractions~\cite{Manohar:2006nz},
clarifies the issue regarding the known importance of the $i\varepsilon$ prescription in obtaining the proper result for the two-loop Coulomb potential~\cite{Peter:1996ig, Schroder:1998vy}.

\section{Review of NRQCD with Single Stage Matching}

In vNRQCD we match onto the EFT at $m$, the mass scale of the heavy quarks. In the low energy theory there are two relevant scales $mv$ (soft) and $mv^2$ (ultrasoft), where $v$ is a scaling parameter that is of order the relative velocity between the heavy particles. The heavy quarks and antiquarks whose dynamics we wish to describe have kinetic energy $p^0\sim mv^2$ and momenta of ${\bf p}\sim mv$. At the $p^\mu\sim mv$ scale there are corresponding soft gluon and massless quark fields, $A_q^\mu$ and $\varphi_q$, and at the $p^\mu\sim mv^2$ scale there are ultrasoft gluon and massless quark fields, $A_{us}^\mu$ and $\varphi_{us}$.  We use dimensional regularization with $d=4-2\epsilon$ to regulate ultraviolet divergences.

At the matching scale $m$ we integrate out all of the hard modes with momenta of order $m$ as well as other offshell modes which include potential gluons with energy $p^0\sim mv^2$ and 3-momentum $\mathbf{p}\sim mv$.  As a result of integrating out offshell modes with momenta $\sim mv$ we generate a set of potentials between the heavy quarks, which are given by spatially non-local four fermion operators. In this one-stage approach there is no further matching at the scale $mv$ since the  $mv^2$ and $mv$ scales are treated simultaneously.

We also choose to integrate out soft heavy quarks with $p^0\sim mv$, which avoids simultaneously having both soft and potential heavy quarks in the EFT.\footnote{This is not strictly necessary. If we instead keep soft heavy quarks $\psi_s$ together with the potential heavy quarks $\psi_{\bf p}$, then the structure of the EFT Lagrangian will be somewhat different. There will be interactions between the soft and potential heavy quarks and soft gluons. The purely soft heavy quark Lagrangian will be identical to HQET, except that we must include 0-bin subtractions to avoid double counting with the potential region. The one-stage matching formalism obtained from this construction would be closer to the formalism of pNRQCD.}  Thus when a soft gluon, with $p^\mu \sim mv$  scatters off a potential heavy quark with $p^\mu \sim (mv^2, m \vec v)$ (relative to its rest mass), the resulting quark is thrown off-shell with $p^0\sim mv$. Therefore, Compton scattering graphs between potential heavy quarks and soft gluons are matched onto non-local (at scales shorter than $1/( mv)$) operators.  These operators essentially act like additional potential interactions between the potential heavy quark and soft gluon and soft light quark modes. We will see that a crucial distinction between NRQCD and the SCET with forward scattering interactions, is that this potential-soft interaction gets corrected beyond tree level in NRQCD, which is not the case for the analogous Glauber exchange operators in SCET~\cite{Rothstein:2016bsq}. We will in fact see that these higher order corrections to this gluon-quark potential are what leads to the new color structure starting at three loops (i.e. the structure that is not ``maximally non-Abelian'' $\propto C_i C_A^k$, where $C_i$ is the tree-level Casmir).

To distinguish the scales $mv$ and $mv^2$, in the EFT we split the external 3-momenta (${\bf P}_{\rm full}$) of the heavy quark fields into a large label piece (${\bf p}\!\sim m v$) and a small residual momentum (${\bf k}\sim mv^2$), so that ${\bf P}_{\rm full}={\bf p} + {\bf k}$. To maintain homogeneous power counting, we use the label formalism developed in \cite{Luke:1999kz}, where momentum or energy of $\sim mv$ are represented by field labels in momentum space. At tree level the relation between the heavy quark fields  in QCD ($\psi$) and vNRQCD ($\psi_{\bf p}$) is
\beq
\psi(x) = \sum_{\bf p} e^{-i {\bf p}\cdot \bf x}\psi_{ {\bf p}}(x)
  \,.
\eeq
An analogous relation holds for the heavy antiquark fields $\chi(x)$ and $\chi_{\bf p}(x)$. 
The full theory derivative operator is then decomposed as
\beq
\label{label}
 i\partial_\mu \rightarrow {\bf P}_\mu+ i\partial_\mu  \,,
\eeq
where the label operator ${\bf P}$~\cite{Bauer:2001ct} only operates on the momentum label space, ${\bf P}^i \psi_{\bf p}(x) = {\bf p}^i \psi_{\bf p}(x)$. 
In this way all derivative operators $\partial_\mu$ acting on fields $\psi_{\bf q}(x)$ will scale as $v^2$.

The effective vNRQCD Lagrangian can be split into soft, ultrasoft, and potential components,
\begin{align}
 \mathcal{L}=\mathcal{L}_{u}+\mathcal{L}_{p}+\mathcal{L}_{s} \,.
\end{align} 
The ultrasoft Lagrangian involves only ultrasoft fields
\beq  \label{eq:Lus}
\mathcal{L}_{u}
  =  \bar\varphi_{us}\, i \slashed{D}\, \varphi_{us} 
   -\frac{1}{4}G^{\mu\nu}_{u}G_{{u},{\mu\nu}} + \ldots
  , 
\eeq
where $G_{u}^{\mu\nu}$ is the ultrasoft field strength and the ellipses are terms that are higher order in $v$. The covariant derivative here is $D^{\mu}=\partial^{\mu}+i\mu^{\epsilon}_{u} \iota^{\epsilon/2} g_{u}(\mu_u) A_u^\mu$, which only contains the ultrasoft gluon field and scales as $D^\mu \sim v^2$,  depends on a coupling at the ultrasoft scale $\mu_{u}=m\nu^2$, where $\nu$ is the subtraction velocity scale. Here $\iota=e^{\gamma_E}/(4\pi)$ appears because the strong coupling is being defined  in the $\overline{\rm MS}$ scheme. The field strength scales as $G_u^{\mu\nu}\sim v^4$.  For convenience we will often suppress the renormalization $Z$ factors that relate bare and renormalized quantities. 

The potential Lagrangian involves both potential heavy quarks and ultrasoft fields and has terms \cite{Luke:1999kz,Grinstein:1997gv, Manohar:1999xd}
\begin{align}
\label{eq:Lp}
\mathcal{L}_{p}
  &=\sum_{ {\bf p}}\Big\{\psi^\dagger_{{\bf p}}\Big[iD^{0}-\frac{({\bf p}-i{\bf D})^2}{2m}
  +\frac{{\bf p}^4}{8m^3} 
  + \frac{c_F g_u}{2m} \mathbf{\sigma}\cdot \mathbf{B}_u 
  \Big]\psi_{ {\bf p}}
  +(\psi \rightarrow \chi)  \Big\}
  \nn\\
  &\quad 
  - \sum_{\bf p,p'}\mu_{s}^{2\epsilon} \iota^\epsilon V({\bf p},{\bf p}^\prime)
 \: \psi^{\dagger}_{{\bf p}^\prime}\psi_{{\bf p}}
    \chi^{\dagger}_{-{\bf p}^\prime}\chi_{-{\bf p}} 
  + {\cal L}_{pu} + \ldots \,,
\end{align}
where ${\cal L}_{pu}$ contains ultrasoft gluon couplings to potential operators and ultrasoft kinetic corrections to the potentials~\cite{Manohar:2000kr}, and the ellipses denote terms of  higher order in the $v$ expansion. The potential heavy quark fields scale as  $\psi_{\bf p}\sim v^{3/2}$ and $\chi_{\bf p}\sim v^{3/2}$, and $\mathbf{B}_u^i = \epsilon^{ijk} G_u^{jk}\sim v^4$. In terms of the velocity subtraction scale the soft scale is $\mu_{s}=m\nu$, and spin and color indices in $V$ and the fermion fields have been suppressed. Matching perturbatively at $m$ and integrating out the potential gluons generates the terms
\begin{align} \label{eq:V}
V({\bf p},{\bf p}^\prime)=
 &(T^{A}\otimes\overline{T}^{A})
  \Big[\frac{{\cal V}_{c}^{(T)}}{{\bf k}^2}
   +\frac{{\cal V}_{k}^{(T)}\pi^2}{m|{\bf k}|}
   +\frac{{\cal V}_{r}^{(T)}({\bf p}^2+{{\bf p}}^{\prime 2})}{2m^2{\bf k}^2}
   +\frac{{\cal V}_{2}^{(T)}}{m^2}
   +\frac{{\cal V}_{s}^{(T)}}{m^2} \mathbf{S}^2 +\frac{{\cal V}^{(T)}_{\Lambda}}{m^2}\Lambda({\bf p}^\prime,{\bf p})
   \nonumber  \\ 
 & +\frac{{\cal V}_{t}^{(T)}}{m^2}T({\bf k})+.... \Big] + (1\otimes1)\Big[\frac{{\cal V}_{c}^{(1)}}{{\bf k}^2}
  +\frac{{\cal V}_{k}^{(1)}\pi^2}{m|{\bf k}|}
  +\frac{{\cal V}_{2}^{(1)}}{m^2} 
  + \frac{{\cal V}_{s}^{(1)}}{m^2}\mathbf{S}^2+....\Big] ,
\end{align}
where ${\bf k}={{\bf p}}^\prime-{\bf p}$, $T^A$ and $\bar T^A$ are color generators in the $3$ and $\bar 3$ representations respectively, and
\begin{align}
& \mathbf{S}= \mathbf{S}_{1}+ \mathbf{S}_{2}, 
& T(\mathbf{k}) &= \mathbf{S}_{1}\cdot  \mathbf{S}_{2}-3\frac{{\bf k}\cdot  \mathbf{S}_{1}{\bf k}\cdot  \mathbf{S}_{2}}{{\bf k}^2}, 
& \Lambda({\bf P}^\prime,{\bf P}) &=\frac{-i\mathbf{S}\cdot ({\bf P}^\prime \times {\bf P})}{{\bf k}^2}
\,.
\end{align}
Here  $\mathbf{S}_{1}$ and $\mathbf{S}_{2}$ are the quark and anti-quark spin operators respectively. The  relation between the bare  and renormalized coefficients is given by ${\cal V}_j^{\rm bare} =\mu_{s}^{2\epsilon} {\cal V}^{\rm R}_j =  {\cal V}^{\rm R}_{j}(\nu)$+ counterterms. These Wilson coefficients were calculated to one-loop in Ref.~\cite{Manohar:2000kr} and the analogous potentials for unequal quark and anti-quark masses were computed at one-loop in Ref.~\cite{Peset:2015vvi}.

\begin{figure}[t]
\centering
\includegraphics[scale=0.6]{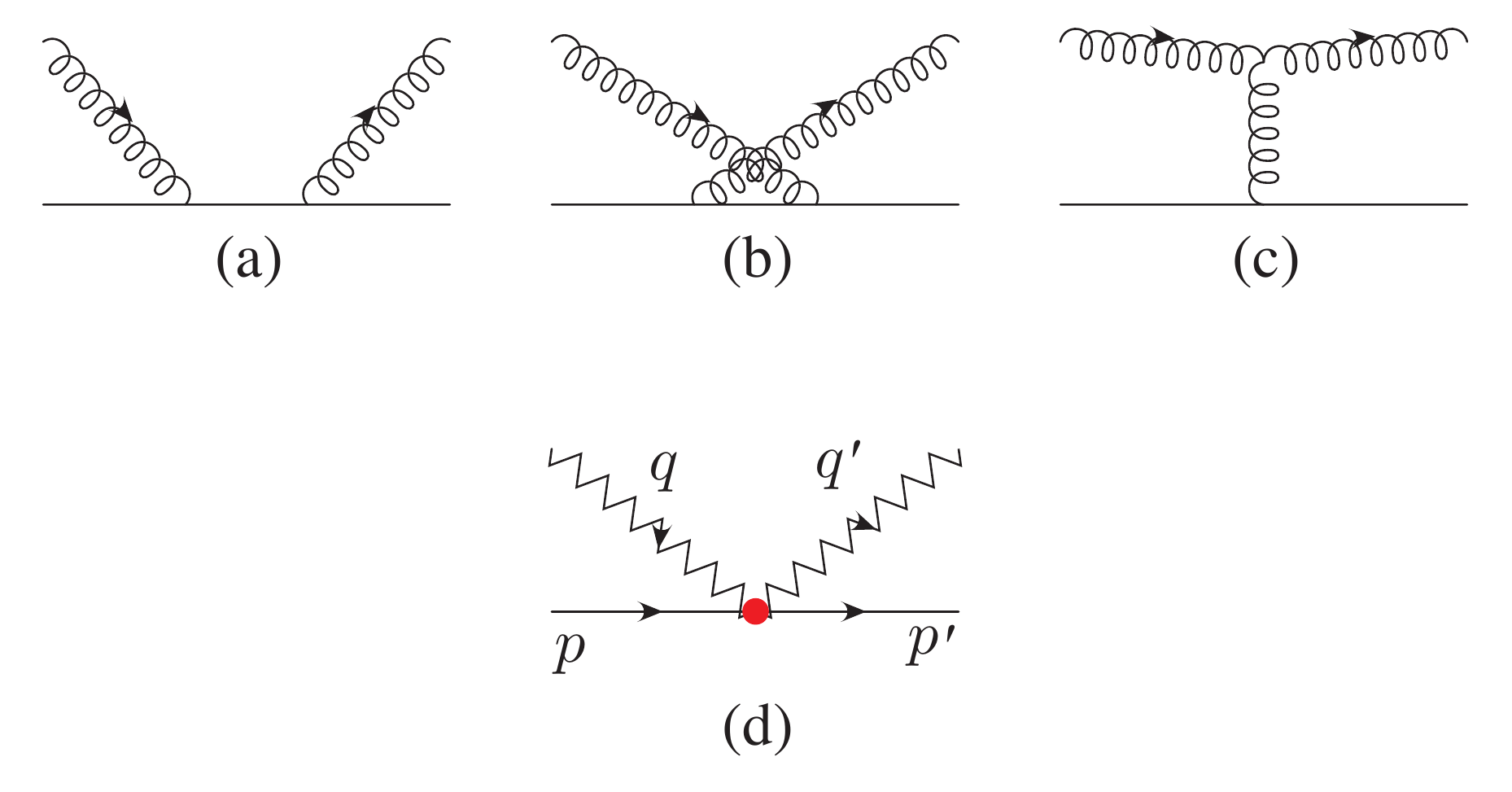}
\caption{The Compton scattering graphs (a, b, c) in QCD match onto the soft gluon coupling (d) in the effective theory at tree level.}
\label{fig:compton}
\end{figure}

The soft Lagrangian consists of pure soft field terms, plus interaction terms involving both soft and potential fields,
\begin{align} \label{eq:Ls}
{\cal L}_s &= 
  \sum_{q,q'} \bigg\{ \bar\varphi_q\: i \slashed{D}^s_{q,q'}\: \varphi_{q'}
  -\frac{1}{4}G_s^{\mu\nu} G^s_{\mu \nu}
  + \frac{1}{\xi_s} \big( q\cdot A_{q} \big)^2
  + \bar c_q (q\cdot i {\cal D}^s_{q,q'})  c_{q'} \bigg\}
  + {\cal L}_s^{\rm int} \,.
\end{align}
We will use the notation $iD_s$ for the soft covariant derivative in position space. In \eq{Ls} the soft covariant derivative is written in momentum space $iD_{q,q'} = \delta_{q,q'}i\partial - g_s(\mu_s) \mu_s^\epsilon \iota^{\epsilon/2} \delta_{q,q'+p} A_p$, and the gauge fixing has been performed in a general covariant gauge with parameter $\xi_s$, and hence includes soft ghosts $c_q$. The soft fields $A_{q}^\mu$, $\varphi_q$, and $c_q$ have incoming momentum $q$, whereas $\bar{\varphi}_q$ and $\bar c_q$ have outgoing momentum $q$. Notice that the gauge fixing of the ultrasoft and soft gluon fields can be performed independently. The ultrasoft fields have momentum components  which are parametrically smaller then their soft counter-parts and therefore these fields do not directly couple to each other. The heavy (potential) fermions do not transform under soft gauge transformation, as such a transformation has energies of order $mv$ and would throw them off-shell.

In the formalism of Ref.~\cite{Luke:1999kz} the soft-potential interaction terms are contained in 
\begin{align} \label{eq:Lsint}
  {\cal L}_s^{\rm int} &=
  - g_s^2 \mu_S^{2\epsilon} \iota^\epsilon \!\!\! 
  \sum_{{\bmp},{\bmp^\prime},q,q^\prime,\sigma} \bigg\{ 
  \frac{1}{2}\, \psi_{\bmp^\prime}^\dagger
  [A^\mu_{q^\prime},A^\nu_{q}] U_{\mu\nu}^{(\sigma)} \psi_{\bmp}
 + \frac{1}{2}\, 
 \psi_{\bmp^\prime} \{A^\mu_{q^\prime},A^\nu_{q}\} W_{\mu\nu}^{(\sigma)}
 \psi_{\bmp}
 \\[2mm]
 &\qquad
  + \, \psi_{\bmp^\prime}^\dagger [\bar c_{q'}, c_q] Y^{(\sigma)}\:
 {\psip p}
 + (\, \psi_{\bmp^\prime}^\dagger T^B Z_\mu^{(\sigma)}\:
 {\psip p} ) \:(\bar \varphi_{q'} \gamma^\mu T^B \varphi_q)  \bigg\}
 + (\psi \to \chi,\: T\to \bar T) \,.\nn 
\end{align}
The notation is such that $\sigma$ denotes a corrections which is down by $v^\sigma$. The leading contribution to this soft Lagrangian arises from the Compton graphs shown in Fig.(\ref{fig:compton}), plus the analogous diagrams with soft quarks and ghosts.  Taking $q_\mu\sim mv$ and $p_\mu\sim (mv^2, mv)$, and keeping only the leading order result in $v$  (which is $\sigma =0$), the result in Feynman gauge is~\cite{Luke:1999kz}
\begin{align} \label{eq:UWYZ0}
 U^{(0)}_{00} &=  \frac{1}{q^0}\,,
 & U^{(0)}_{0i} &= -\frac{{\mathbf (2 p'-2 p-q)}^i}{\ppp}\,, 
 & U^{(0)}_{i0} &= -\frac{{\mathbf (p-p'-q)}^i}{\ppp}\,, 
 & U^{(0)}_{ij} &= \frac{(-\delta^{ij}) 2 q^0  }{\ppp} 
  \,, \nn\\
 W^{(0)}_{\mu\nu} &= 0 \,,
 & Y^{(0)} &=  \frac{-q^0}{\ppp} \,,
 & Z^{(0)}_0 &= \frac{1}{\ppp} \,,
 &  Z^{(0)}_i &=0 \,.
\end{align}
Here the ghost field, $c_{s}$, is the same one used in \eq{Ls}. The normalization for the generators in the fundamental representation have been taken to be  ${\rm Tr}(T^a T^b)=\frac{1}{2}\delta^{ab}$. The terms that are higher order in the velocity with $\sigma=1,2$ have also been calculated in Refs.~\cite{Manohar:1999xd,Manohar:2000hj}. The soft-potential interactions are responsible for part of the running of the Wilson coefficients potential, $V({\bf p}, {\bf p}^\prime)$ in \eq{Lp}, and they are solely responsible for the running of the leading order Coulomb potential operator until 3-loops~\cite{Luke:1999kz,Hoang:2001rr,Hoang:2002yy}. 

Note that in the Abelian case the sum of the two Compton scattering graphs cancel, and hence $W_{\mu\nu}^{(0)}=0$. This cancellation would not arise if we kept the $i \varepsilon$ in the offshell soft fermion propagator. Keeping this $i \varepsilon$ would mean including a contribution where the gluons are immediately forced to be potential and not soft. 
To see this write the propagator with a principal value and $\delta$-function as
\begin{equation}  \label{eq:epsilon}
  \frac{1}{q_0+i\varepsilon}=P(\frac{1}{q_0})-i \pi \delta(q_0).
\end{equation}
The delta function forces the gluon to have vanishing energy, i.e. it becomes a potential gluon. 
Thus in order not to double count we should not include this twice. The rigorous procedure which enforces this  are the vNRQCD ``zero-bin" subtractions formulated in~\cite{Manohar:2006nz}, which avoids double counting between soft and potential contributions. This will be discussed further below.

We emphasize that the individual results in \eq{UWYZ0} for $\sigma=0$ and their analogs for $\sigma>0$  are gauge {\it dependent}. As previously mentioned, there is nothing wrong in working this way, as the $S$-matrix elements will always be gauge invariant. However, by not keeping manifest soft gauge invariance we lose a useful organizing principle and there is a corresponding loss of simplicity. The main goal of this work is to setup a formalism which will make the coefficients in the Lagrangian ${\cal L}_s^{\rm int}$ manifestly soft gauge invariant.

\section{Reformulation of the Soft NRQCD Lagrangian}

The gauge non-invariance of the Lagrangian ${\cal L}_s^{\rm int}$ arises because even though we have used the onshell conditions $k^2=0$ for the external soft gluons and analogs for the heavy fermions, we have not used our complete freedom to exploit the on-shell transversality condition $\partial \cdot A=0$. In this section we will construct soft gauge invariant operators that match the full theory amplitudes on-shell when exploiting the full freedom of the equations of motion.  This same organization was used in Ref.~\cite{Rothstein:2016bsq} in constructing gauge invariant Glauber exchange operators in SCET.

First we introduce the soft gluon and soft quark gauge invariant building blocks, which in position space read\footnote{Our sign convention for $g$ with $iD^\mu = i\partial^\mu - gA^\mu$ is opposite to the one used in the SCET literature, but agrees with the convention used in vNRQCD.}
\begin{align} \label{eq:Bdefn}
  B^\mu(x) &= - \frac{1}{g_s}S^{\dagger}_{\rm  v}(x,-\infty)\, iD_s^\mu (x)\, 
  S_{\rm  v}(x,-\infty)
  \,, \\
\Xi(x) &= S^{\dagger}_{\rm  v}(x,-\infty)\, \varphi(x)
  \,,\nn
\end{align}
where $S_{\rm  v}$ is a soft Wilson line in the ${\rm  v}^\mu$ direction,  which parallel transports the derivative to time-like $-\infty$. Here ${\rm v}=(1,\vec 0)$ in the heavy $Q\bar Q$ rest frame (we use a roman v to make it clear that this is not the relative velocity scaling parameter $v$). In particular the fundamental soft Wilson line is
\begin{align}   \label{Sexp}
 S_{\rm  v}(x,-\infty) 
 &= P \exp\bigg( - ig_s \int_{-\infty}^0 \!\!\! d\lambda \: 
 {\rm v}\cdot A(\lambda {\rm v} + x) \bigg)
 =  1-ig_s\int_{-\infty}^0 \!\!\! d\lambda\ A^0(\lambda {\rm v}+x)  + \ldots
  \nn \\
 &= 1-ig_s\int_{-\infty}^0 d\lambda
  \sum_{ k} e^{-i\lambda k^0 -i x\cdot k} A^0_k + \ldots
  \nn \\
 &= 1+ g_s\sum_{ {k}}\frac{ A^0_k }{ {k}_0 +i\varepsilon} e^{-i { {k}}\cdot x}
  +\ldots \,,
\end{align}
and we note that $S_{\rm v}^\dagger(x,-\infty)=S_{\rm v}(-\infty,x)$. 
The equations of motion ${\rm v}\cdot D_s\, S_{\rm v}=0$ obeyed by the soft Wilson line imply that ${\rm v}\cdot B=0$. The $i\varepsilon$ prescription in the denominators is determined by the limits of integration, and kills the contribution from the endpoint at infinity. Note that we used the same Wilson line on both sides of the covariant derivative in the definition of $B^\mu$. This ensures that $B^\mu(x)$ is a pure octet. In particular using standard relations between fundamental and octet Wilson lines and \eq{Bdefn} gives
\begin{align} \label{eq:Boctet}
  B_\mu = \frac{1}{{\rm v}\cdot\partial_s} {\rm v}^\nu G^{s\,b}_{\nu\mu} {\cal S}_{\rm v}^{ba} T^a  \equiv B_\mu^{ a} T^a \,.
\end{align}
This shows that $B^{a}_{\mu}$ is a soft field strength attached to an adjoint Wilson line ${\cal S}_{\rm v}={\cal S}_{\rm v}(x,-\infty)$. 
The building blocks have scaling $B^\mu\sim v$ and $\Xi(x)\sim v^{3/2}$. They both start off linear in the soft gluon or quark fields, respectively, for which they are soft gauge invariant extensions. In momentum (label) space using Eq.~(\ref{Sexp}) leads to
\begin{align}  \label{eq:Bexpn}
B_k^{\mu} &=A^{\mu}_k-\frac{k^{\mu}A^0_k}{k^0} 
   +g_s \biggl\{\Bigl[A^{\mu}_{k_{1}},\frac{A^{0}_{k_{2}}}{k_{2}^{0}}\Bigr] -\frac{k_{1}^\mu}{(k_{1}^0+k_{2}^0)} 
  \Bigl[A^{0}_{k_{1}},\frac{A^{0}_{k_{2}}}{k_{2}^{0}}\Bigr] \bigg\}
 + \ldots
 \nn \,, \\
 \Xi_k &= \varphi_k  - g_s \frac{A_{k_1}^0}{k_1^0} \varphi_{k_2} + \ldots
   \,,
\end{align}
where here $k_1+k_2=k$, and all denominators have $k_i^0 = k_i^0+i\varepsilon$ with our convention for the soft Wilson lines.  Note that the operators $B^\mu$ and $\chi_s$ transform under a global gauge transformation (equivalent to its transformation at $-\infty$), but this transformation will be canceled by other fields in the interaction Lagrangian we will construct below.

The convention for the direction of the Wilson lines in \eq{Bdefn}, or equivalently the sign of the $i\varepsilon$ in \eq{Bexpn}, should be irrelevant since the pole only matters  in the potential region as described in \eq{epsilon}, and hence is not a part of the soft sector of the theory\footnote{Similar reasoning in the EFT of black hole binaries \cite{Goldberger:2004jt} explains apparent IR divergences in the calculation of gravitational potentials at $O(v^8)$\cite{Porto:2017dgs}.}. Therefore, the direction of the Wilson line  must be irrelevant if we perform the proper potential zero-bin subtraction \cite{Rothstein:2016bsq}. To demonstrate this we  consider an alternate definition for the soft gluon building block operator
\begin{align} \label{eq:Bdefnalt}
(B^\mu)^{\rm alt} 
  &= -\frac{1}{g}S_{\rm v}^\dagger(x,\infty) iD_s^\mu (x) S_{\rm v}(x,\infty)
   \,.
\end{align}
This definition still leads to an octet $B^\mu$ field, but it now gives a field strength attached to an adjoint Wilson line ${\cal S}_{\rm v} = {\cal S}_{\rm v}(x,\infty)$.  Unless otherwise specified we will by default work with the definition in \eq{Bdefn}.

At lowest order in the $v$ power counting we now consider building operators for the interaction Lagrangian ${\cal L}_s^{\rm int}$ with two potential heavy-quark fields and two soft gluon building blocks $B^\mu$ (which is the minimum number allowed that is consistent with soft momentum conservation), $\psi_{   {\bf p}^\prime}^\dagger  B^\mu_{q} B^\nu_{q^\prime} \psi_ {  {\bf p}}$.\footnote{One may wonder why there is no such operator in HQET given that in the one particle sector there should be no distinction between NRQCD and HQET. However, we must keep in mind that  these operators will be inserted into time ordered products and thus the relevant dispersion relation for the quarks will always be that of NRQCD.  In HQET the Compton scattering is reproduced by a time ordered product whereas in NRQCD we use these 
higher dimensional, leading order in $v$, operators with the soft heavy fermion integrated out.}  This product of fields is ${\cal O}(v^5)$, and according to the power counting theorem of Ref.~\cite{Luke:1999kz} it must be multiplied by a coefficient 
$C(q,q^\prime, {\bf p}, {\bf p}^\prime)\sim v^{-1}$ to give a leading power contribution. (Although this will make it leading order in $v$, the interaction also starts at ${\cal O}(\alpha_s)$, and hence is perturbative if $\mu_s\gg\Lambda_{\rm QCD}$.)  Imposing the fact that the leading power coefficient must also be $m$ independent, and can not be more singular than linear or quadratic in a momentum, due to the locality of QCD,\footnote{Here we are restricting ourselves to two to two scattering of point particles.} we can eliminate the possibility of contracting the indices $\mu$ and $\nu$ with momenta, and must therefore have a $g_{\mu\nu}$ at leading power. There are three independent ways to construct a color singlet out of  $3$, $\bar 3$, and two $8$s, given by the following tensors $if^{abc} T^c_{\alpha\bar\beta}$, $d^{abc} T^c_{\alpha\bar\beta}$, $\delta^{ab} \delta_{\alpha\bar\beta}$. This gives
\begin{align} \label{eq:Lsintnew}
  {\cal L}_s^{\rm int} &=
  - g_s^2 \mu_S^{2\epsilon} \iota^\epsilon\!\!\! 
  \sum_{{\bmp},{\bmp^\prime},q,q^\prime,\sigma} \bigg\{ 
  \frac{i}{2}f^{abc}\,U_{ij}^{(\sigma)} (\psi_{\bmp^\prime}^\dagger T^c  \psi_{\bmp})(
  B^{i,a}_{q^\prime}B^{j,b}_{q} )
 + \frac{1}{2}d^{abc}\, W_{ij}^{(\sigma)}
( \psi_{\bmp^\prime} T^c \psi_{\bmp})
 (B^{i,a}_{q^\prime}B^{j,b}_{q})
 \nn \\
 &\qquad
 +\frac{1}{2} R_{ij}^{(\sigma)} \delta^{ab}
( \psi_{\bmp^\prime}^\dagger \psi_{\bmp})
 (B^{i,a}_{q^\prime}B^{j,b}_{q})
 + ( \psipd{p^\prime} T^B Z_\mu^{(\sigma)}
 {\psip p} ) \:(\bar \Xi_{q'} \gamma^\mu T^B \Xi_q)
 + ( \psipd{p^\prime}  Z_\mu^{\prime(\sigma)}
 {\psip p} ) \:(\bar \Xi_{q'} \gamma^\mu  \Xi_q)  \bigg\}
 \nn\\
 &\qquad 
  + (\psi \to \chi,\: T\to \bar T) \,.
\end{align}
Here $U_{ij}^{(\sigma)}$, $W_{ij}^{(\sigma)}$, $R_{ij}^{(\sigma)}$, $Z_{\mu}^{(\sigma)}$, and $Z_{\mu}^{\prime (\sigma)}$ are functions of $({\bf p,p'},q)$ with $q'={\bf p'-p}-q$, and $g_s=g_s(\mu_s)$ as before. Note that only spatial $B^{i}$s appear since ${\rm v}\cdot B=0$. 
It is convenient to include the factor of $1/2$ since the Feynman rule will include terms where the contraction with $B_{q'}$ is swapped with those for $B_q$, inducing a factor of 2. 
The kinematics are such that  all of the ${\cal O}(v)$ energy has to flow through the soft gluons or quarks in order for the initial and final state heavy quark to remain close to on shell with energy of ${\cal O}(v^2)$.

With \eq{Lsintnew} we have a manifestly soft and ultrasoft gauge invariant form  for the interactions between potential heavy quarks and soft fields. Under a soft gauge transformation $B_q^i$ and $\Xi_q$ are invariant, and $\psi_{\bf p}$ and $\chi_{\bf p}$ do not transform, so the Lagrangian is invariant. Under an ultrasoft gauge transformation $\psi_{\bf p}$ and $\chi_{\bf p}$ have the usual fundamental matter field transformations, and the soft fields $B_q^i$ and $\Xi_q$ also transform like adjoint and fundamental matter fields since only the global color charge is seen by these long wavelength gauge transformations. Therefore the Lagrangian is also ultrasoft gauge invariant. 

\subsection{Leading Power Soft-Potential Interactions}

Let us now match the leading order operator to obtain $U^{(0)}_{ij}$. Its value is fixed so as to reproduce the Compton scattering tree level graphs  in Fig.{\ref{fig:compton}}. Consider first matching only the spatial polarizations for both gluons, for which the  only a contribution stems from the diagram with a 3-gluon vertex. We find
\begin{align} \label{eq:WC}
 U_{ij}^{(0)}(q,q^\prime, {\bf p}, {\bf p}^\prime)  =  -\frac{2 q^0 \delta_{ij}}{\left( {\bf{p}}^\prime-{\bf {p}} \right)^2} \,,
\end{align}
which agrees with the result for $U^{(0)}_{ij}$ from \eq{UWYZ0}. 
Naively it would appear that with this fixed value of the coefficient, we will not match the time-like polarizations. However, gauge invariance implies that
on-shell the results for the time-like polarizations must also match. To check this explicitly, we compute the tree level full QCD amplitude involving time-like polarizations from the graphs in \fig{compton}, and write it back in the form of soft gluon fields $\epsilon_q^{\mu A} T^A \to A_q^{\mu A} T^A = A_q^\mu$ to obtain 
\begin{align} \label{eq:Mfull}
M_{\rm full}
&= \biggl\{ \frac{1}{q^0} 
 - \frac{2q_0}{ \left( {{\bf{p}}}^\prime \!-\!  {\bf {p}} \right)^2}
 \biggr\} [A^0_{q},A^0_{q^\prime}] 
  + \frac{ {2\bf q}^i}{ \left( { {\bf{p}}}^\prime\!-\! {\bf {p}} \right)^2} [A_{q}^0, A_{q^\prime}^i] 
 - \frac{ {2\bf q}^{\prime\,i} }{ \left( {{\bf{p}}}^\prime\!-\!  {\bf {p}} \right)^2}  [A_{q}^i, A_{q^\prime}^0] 
 \,,
\end{align}
as the structure appearing in $i g^2 \psi_{\bf p'}^\dagger M_{\rm full} \psi_{\bf p}$ when expanding the amplitude to leading order in $v$. 
To derive this result we have made use of momentum conservation and the equations of motion $q^{0}A_{q}^{0}= {\bf q}\cdot {\bf A}_{q}$. To obtain the analog of $M_{\rm full}$ in the effective theory we simply replace each of $B_q$ and $B_{q'}$ in \eq{WC} by the terms with one gluon field from \eq{Bexpn} to obtain the terms involving at least one time-like polarization, again substituting $\epsilon_q^{\mu A} T^A \to A_q^\mu$,
\begin{align} \label{eq:Mvnrqcd}
 M_{\rm vNRQCD} 
  &= 2\Big(\frac{{\bf q\cdot q'}}{q^0\left( { {\bf{p}}}^\prime\!-\!   {\bf {p}} \right)^2 } [A^0_{q},A^0_{q^\prime}] 
  + \frac{ {\bf q}^i}{ \left( { {\bf{p}}}^\prime\!-\! {\bf {p}} \right)^2} [A_{q}^0, A_{q^\prime}^i] 
 - \frac{ {\bf q}^{\prime\,i} }{ \left( { {\bf{p}}}^\prime\!-\!   {\bf {p}} \right)^2}  [A_{q}^i, A_{q^\prime}^0] \Big)
\end{align}
Using the onshell conditions $q^2=q^{\prime\,2}=0$ and momentum conservation implies that ${\bf q}\cdot {\bf q^\prime}= \frac{1}{2}(  {\bf{p}}^\prime-  {\bf {p}} )^2 -(q^0)^2$, and hence that \eq{Mvnrqcd} matches exactly with \eq{Mfull} onshell. 

From tree level matching the  remaining leading power coefficients are given by 
\begin{equation} \label{eq:UWZ0}
 R_{ij}^{(0)}=W^{(0)}_{ij} = 0, \qquad
  Z^{(0)}_0 = \frac{1}{\ppp}, \qquad
   Z^{(0)}_i =Z^{\prime (0)}_\mu =0 \,.
\end{equation}
Note that since we are matching on-shell we no longer have any analog of the operator in \eq{Lsint} that involves an interaction of soft ghosts with potential heavy quarks.  \eq{Lsintnew} alone reproduces the correct gauge invariant anomalous dimension results as discussed further in the next section. 

\begin{figure}[t]
\centering
\includegraphics[scale=0.6]{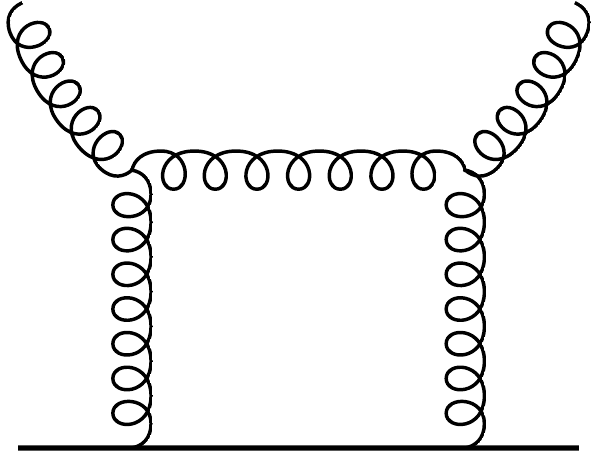} 
\hspace{0.4cm}
\raisebox{-0.07cm}{\includegraphics[scale=0.6]{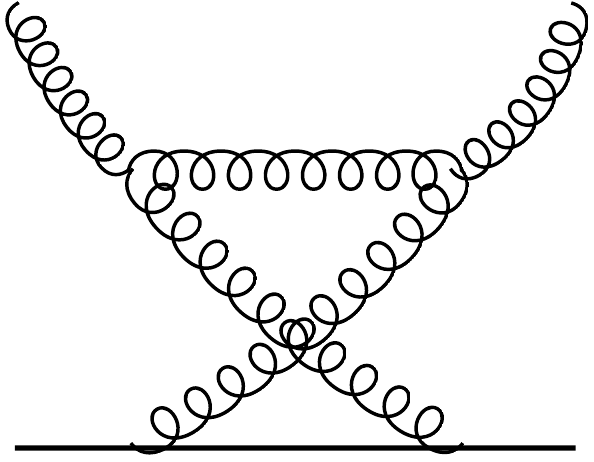}} 
\hspace{0.4cm}
\raisebox{-0.02cm}{\includegraphics[scale=0.6]{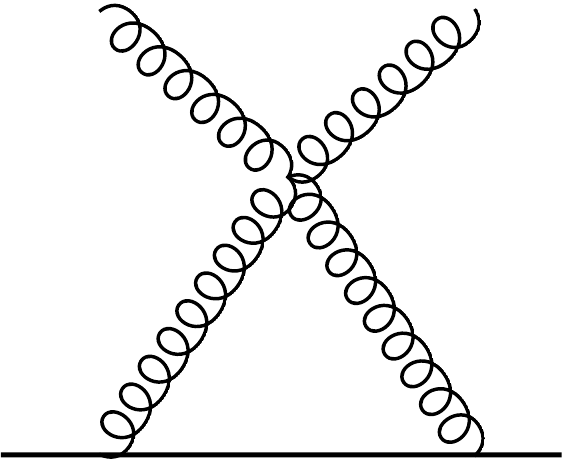}}
\caption{Full theory one loop graphs for matching onto the operators with $d^{abc} B^{ia}B^{jb}$ or $\delta^{ab} B^{ia}B^{jb}$ in \eq{Lsintnew}.}
\label{fig:full}
\end{figure}

While at tree level  $W_{ij}^{(0)}$ and $R_{ij}^{(0)}$ vanish,  these coefficients will be generated by the matching at one loop.   
In the full theory we have the three one loop diagrams shown in \fig{full} which can contribute to the $d^{abc}$ and $\delta^{ab}$ color structures,  while there are no one-loop vNRQCD diagrams with these color structures to subtract. 
These diagrams generate non-zero results $W_{ij}^{(0)}={\cal O}(\alpha_s)$ and $R_{ij}^{(0)}={\cal O}(\alpha_s)$. 
In the language of the threshold expansion~\cite{Beneke:1997zp}, these one loop matching results come from the potential region of the full theory loop integrals.  In vNRQCD these contributions generate potentials for the soft gluons as in \eq{Lsintnew}. There will also be one loop corrections to $U_{ij}^{(0)}$ that require the calculation of additional full theory diagrams and the subtraction of vNRQCD one loop diagrams. 
Interestingly, the time-ordered product of the $W_{ij}^{(0)}$ one-loop vertices yields three loop corrections to the static potential which contribute to the non-trivial $d_A^{abcd} d_F^{abcd}$ color structure~\cite{Smirnov:2008fk,Smirnov:2009uq,Anzai:2009tm, Smirnov:2010vn}.

\begin{figure}[t]
\centering
\includegraphics[scale=0.6]{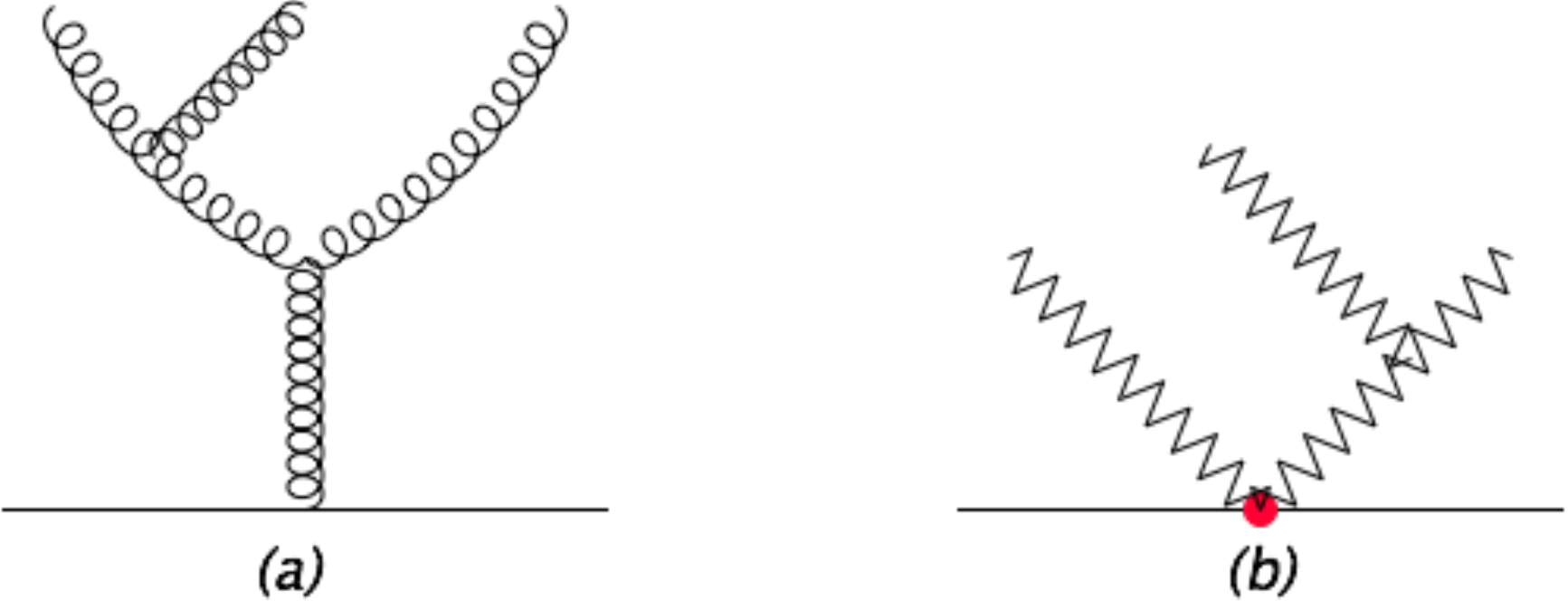}
\caption{Graphs to consider for matching onto operators with 3 $B^i$s. Graph (a) is a QCD graph that exists at leading power with three spatial polarizations, while (b) is a time-ordered product graph in vNRQCD.}
\label{fig:compton3}
\end{figure}
We may also investigate operators that have more building block fields. For example, we may consider operators with three $B$'s, $\psi_{   {\bf p}^\prime}^\dagger  B^i_{q} B^j_{q^\prime} B^k_{q^{\prime\prime}} \psi_ {  {\bf p}}$ with all possible color structures.
It is straightforward to show that the matching onto these operators vanishes at tree level. To match we can consider full theory graphs which are leading power and have three spatial polarizations for the gluons.
Since the attachment to the heavy quark has a time-like polarization in the full theory at leading power, the only such diagram is the one shown in \fig{compton3}(a).  The result for this graph is exactly reproduced on-shell by the vNRQCD graph in \fig{compton3}(b) with all spatial polarizations for the gluons.
Thus the tree level matching onto the three $B$ operator leads to a vanishing Wilson coefficient. Examining operators with more spatial gluons does not change this tree level correspondence between QCD and vNRQCD graphs when we have a single gluon attached to the heavy quark, generalizing the graphs in Fig.~\ref{fig:compton3}. This remains true even when there is more than one gluon attached to the heavy quark.
Thus  all operators of the form $\psi_{   {\bf p}^\prime}^\dagger B^{i_1}_{q_1}\cdots B^{i_n}_{q_n} \psi_ {  {\bf p}}$ with $n\ge 3$ have vanishing Wilson coefficients at tree level. When matching at higher loop orders the coefficients of these operators may no longer vanish.

\subsection{Power Corrections to Soft-Potential Interactions}

The above analysis can be extended to use the gauge invariant soft building blocks to include terms at subleading orders in the $v$ expansion in Eq.(\ref{eq:Lsintnew}). The Wilson lines in the $B_q^\mu$ and $\Xi_q$ contain all the coupling to the time-like gauge boson polarizations, including terms like $U_{00}^{(\sigma)}$ and $U_{0i}^{(\sigma)}$ in \eq{UWYZ0}. Therefore  we may extract  the required subleading power matching results from~\cite{Manohar:2000hj, Manohar:1999xd} by keeping only the results with spatial indices, which to $1/m^2$ read
\begin{align} \label{eq:UWZ2}
U^{(1)}_{ij} 
 &= i c_{F}\frac{\epsilon_{ijk}{\bsigma}_{k}}{2m}
  +[2\delta_{ij}{\bf q}_{m}+\delta_{im}(2{\bf p}^\prime-2{\bf p}-{\bf q})_{j}+\delta_{jm}({\bf p}-{\bf p^\prime}-{\bf q})_{i}]
  \\*
 &\quad \times\Big[\frac{({\bf p}+{\bf p^\prime})_{m}+i c_{F} \epsilon_{mkl}({\bf p}-{\bf p^\prime})_{k}{\bsigma}_{l}}{2m({\bf p}-{\bf p^\prime})^{2}}\Big] 
  \,, \nonumber \\* 
U^{(2)}_{ij}
 &= \frac{({\bf p}+{\bf p}^\prime)_{i}({\bf p}+{\bf p}^\prime)_{j}}{4m^2q^{0}}
  +\frac{c_{F}^2({\bf p}-{\bf p}^\prime)\cdot {\bf q}\,\delta_{ij}}{4m^2q^0}
  +\frac{ic_F({\bf p}+{\bf p}^\prime)_{j}[({\bf p}-{\bf p}^\prime)\times{ \bsigma}]_{i}}{4 m^2 q^0} 
 \nonumber \\ 
 & - \frac{i c_F \epsilon_{ijk}{\bf q}_{k}({\bf p}+{\bf p}^\prime)\cdot{ \bsigma}}{4m^2q^0}
 + \frac{i c_F \epsilon_{ijk}{\bsigma}_{k}({\bf p}+{\bf p}^\prime)\cdot{\bf q}}{4m^2q^0}+\frac{(1-c_F^2){\bf q}_i({\bf p}-{\bf p}^\prime+{\bf q})_{j}}{4m^2q^0} 
 \nonumber \\ 
 & +\frac{c_F^2{\bf q}^2\delta_{ij}}{4m^2q^0}-\frac{i\delta_{ij}q^0c_S\, i{\bsigma}\cdot({\bf p}^\prime\times{\bf p})}{2m^2({\bf p}^\prime-{\bf p})^2} 
 - \frac{2 q^0 \delta_{ij} ({\bf p'}\,^2-{\bf p}^2)^2 }{ 4 m^2 ({\bf p'-p})^4 }
  \,, \nonumber \\ 
W^{(1)}_{ij}
 &=\frac{\delta_{ij}}{2m} 
  \,,\nn\\
 Z^{(1)}_0 &= 0 \,, \quad
 Z^{(1)}_i =  \frac{ -(\mathbf{{p+p'}})^i - i c_F [(\mathbf{{p-p'})\times
 \bsigma]}^i}{2m \ppp } 
\,,\quad \nn\\
 Z^{(2)}_0 &= -\frac{1}{4 m^2} \bigg[ \frac{c_D}{2} - \frac{c_S\, \spp}
   {\ppp} \bigg] \,,\quad
 Z^{(2)}_i = 0 \,.  \nn
\end{align}
This provides a much simpler set of matching coefficients than the full list of terms including direct $A^0$ couplings.
Here $c_F(\mu_S)$, $c_S(\mu_S)$, and $c_D(\mu_S)$ are Wilson coefficients in HQET Lagrangian whose renormalized low scale values can be found in~\cite{Manohar:1997qy} and whose leading logarithmic RGE equations can be found in~\cite{Bauer:1997gs}.

 \section{One Loop Running}

\subsection{Running of the Leading Power Potential}

To test some aspects of our gauge invariant formulation of ${\cal L}_s^{\rm int}$ we consider in this section the soft one-loop diagrams that are responsible for the running of the Coulomb potential. We also discuss the independence of results to the direction chosen for the Wilson lines in $B^\mu$.

\begin{figure}[t]
	\centering
	\includegraphics[scale=0.25]{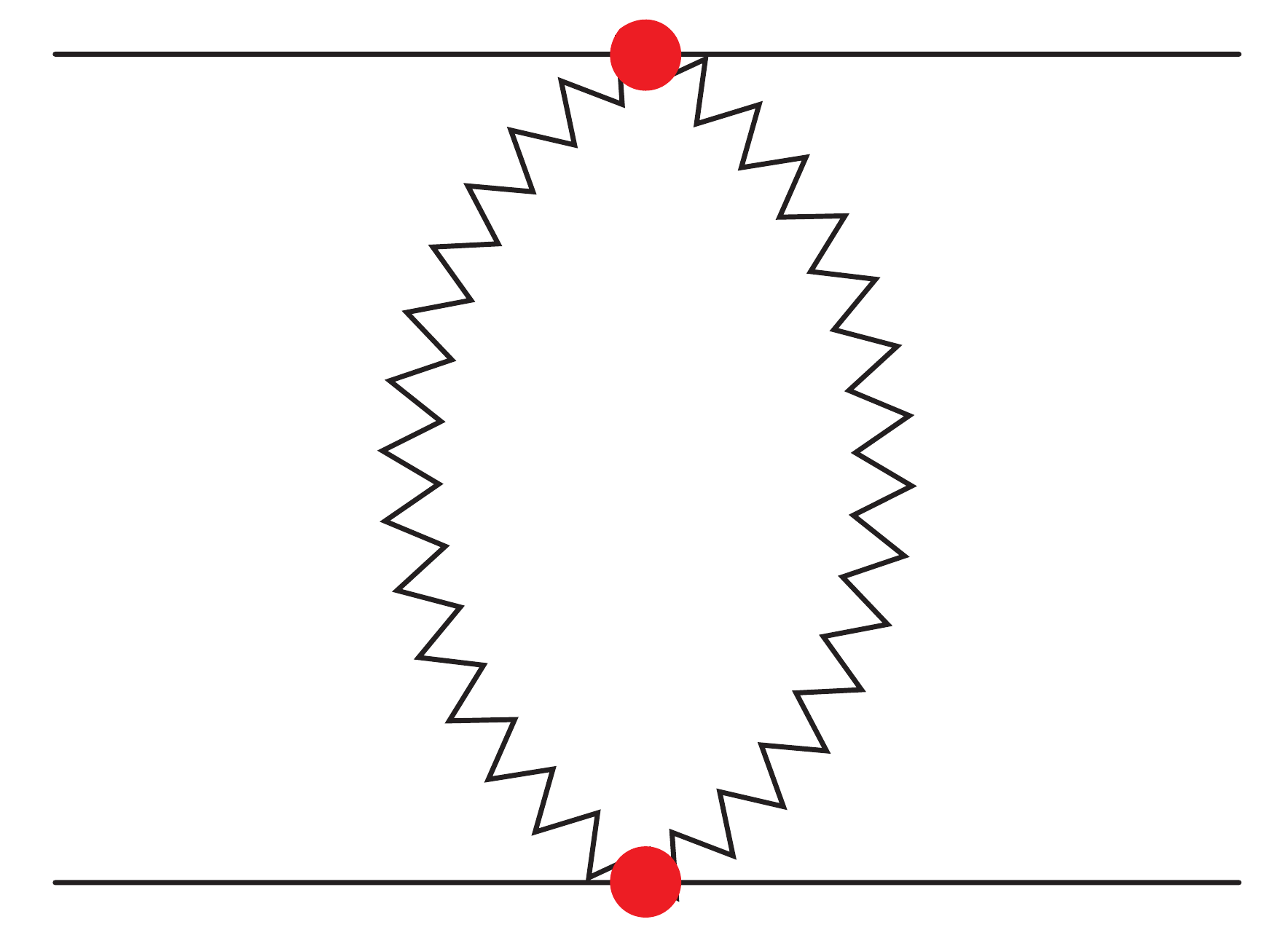}
	\caption{Scattering of heavy quarks at one loop through a soft time-ordered product diagram with soft gluons or soft quarks in the loop using the Lagrangian in \eq{Lsintnew}.}
	\label{fig:oneloop}
\end{figure}

The one loop diagram shown in \fig{oneloop} arises from  time ordered product of two $U_{ij}^{(0)}$ operators  and is responsible for leading log running of the Coulomb potential coefficient ${\cal V}_c^{(T)}$. Since the soft-fermion loop calculation is identical to that considered in earlier papers (see for example Ref.~\cite{Luke:1999kz}), we will only consider the contributions from soft gluons. This calculation can be simplified by considering the contraction of the full $B^\mu$ field, rather than the soft $A_s$ fields it contains (where $A_s$ is still the quantized field).  Since $B^0=0$, we only need the time ordered product of two spatial $B^i$ fields. To leading order in $g$ this is given by the gauge invariant result  
\begin{align}
\label{eq:BBTOP}
\int d^4x \, e^{iq\cdot x} 
  \big\langle 0 \big| T B^{iA}(x) B^{jB}(0) \big| 0 \big\rangle
 &= 
\int d^4k\:   \bigl\langle 0 \big| B_q^{iA} B_{k}^{jB}
  \big| 0 \bigr\rangle
  \\ 
 & =\frac{i\, \delta^{AB}}{q^{2}+i\varepsilon} \bigg[\delta^{ij}- \frac{q^i q^j}{(q^0+i\varepsilon)(q^0-i\varepsilon)} \bigg] + O(g^2) 
  \,. \nn
\end{align}
Here the labels $q$ and $k$ on the $B$s are continuous variables.
It is easy to check that this same result is obtained using any general covariant gauge choice with gauge parameter $\xi_s$.  Here the prescription for the sign of the $i\varepsilon$ in the static lines  $q_0\pm i\varepsilon$ is fixed by the Wilson line directions in \eq{Bdefn}. Both signs appear in \eq{BBTOP} because for one $B$ the momentum $q$ is incoming, while for the other $B$ the momentum $q$ is outgoing. The result in \eq{BBTOP} also does not depend on whether we start with the definition of $B$ in \eq{Bdefn} or the alternate in \eq{Bdefnalt}.  We also note that, not unexpectedly, the propagator result on the right-hand-side is equivalent to that for the soft gluon in an $A^0=0$ gauge, up to the issue of how the $q^0$ poles are handled. (In our analysis zero-bin subtractions play a key role in handling these poles.)

Using dimensional regularization with $d=4-2\epsilon$ for the graph in \fig{oneloop} we find 
\begin{equation}  \label{softTproduct}
 {\rm Fig.}\ref{fig:oneloop}
  =\frac{4g_{s}^4 \mu_{s}^{2\epsilon} \iota^{\epsilon} C_A}{2 {\bf k}^4} 
 \, (T^{A}\otimes\overline{T}^{A})\ \mu_{s}^{2\epsilon} \iota^{\epsilon}
  \!\! \int\!\! \frac{d^dq}{(2\pi)^d} \: T_s(q,{\bf k})
  ,
\end{equation}
where we have included a symmetry factor of $1/2$, and the $\bar T_A$ is for the antiquark which is in the $\overline{3}$ representation.  Here ${\bf k}={\bf p'-p}$ is the momentum transfer and the loop integrand is
\begin{align} \label{eq:Tsqk}
  T_s(q,{\bf k}) &= \frac{(q^0)^2}{[q^2+i\varepsilon][(q+{\bf k})^2+i\varepsilon] } 
 \bigg[\delta^{ij}-\frac{q^i q^j}{(q^0\!+\!i\varepsilon)(q^0\!-\!i\varepsilon)} \bigg]
 \bigg[\delta^{ij}-\frac{(q+k)^i (q+k)^j}  
    {(q^0\!+\!i\varepsilon)(q^0\!-\!i\varepsilon)} \bigg]
   \\
 &= \frac{(d-2)(q^0)^2  - {\bf k}^2}{[q^2+i\varepsilon][(q+{\bf k})^2+i\varepsilon]}
  + \frac{\big[ \frac{q^4}{2} + q^2 {\bf k}^2 +\frac{{\bf k}^4}{4} \big]}
    {[q^2+i\varepsilon][(q+{\bf k})^2+i\varepsilon]} \frac{(q^0)^2}{(q^0\!+\!i\varepsilon)^2(q^0\!-\!i\varepsilon)^2}
   \,. \nn
\end{align}
To go from the first to second line of \eq{Tsqk} we have written all terms in the numerator in terms of either propagator denominators, or $q^2$, $(q^0)^2$, or ${\bf k}^2$, and dropped power law divergent scaleless integrals. The first term in the integrand in \eq{Tsqk} gives a standard integral,
\begin{align} \label{eq:softintegral1}
  I_1 &= \mu_{s}^{2\epsilon} \iota^{\epsilon} \!\!
   \int\!\! \frac{d^{d}q}{(2\pi)^d} \frac{ (d-2)(q^0)^2 -{\bf k}^2  }{[q^2+i\varepsilon][ (q+{\bf k})^{2}+i\varepsilon]} 
  = \frac{i{\bf k}^2}{16\pi^2} \bigg( \frac{5}{6}\big(-\frac{1}{\epsilon_{\text{uv}}}
  - \ln \frac{\mu_{s}^2}{{\bf k}^2}\big) - \frac{31}{18} \bigg) 
  \,. 
\end{align}
For the second term in the integrand in \eq{Tsqk} we have a pinch singularity from the $q^0\pm i\varepsilon$ denominators, as can be verified by working out terms in the residue theorem for the $q^0$ contour integral.  However this pinch occurs from the potential region $q^0\approx 0$ where the approximations used to derive the soft Lagrangian are not valid. In the vNRQCD theory the contribution is removed from the soft integrand by a zero-bin subtraction~\cite{Manohar:2006nz} that is intrinsic to the definition of the soft gluon propagators.  Defining $T_{s2}(q_0^2,{\bf q},{\bf k}) = \big[ \frac{q^4}{2} + q^2 {\bf k}^2 +\frac{{\bf k}^4}{4} \big]/\big\{
{[q^2+i\varepsilon][(q+{\bf k})^2+i\varepsilon]}\big\}$ the second term in the integrand with the potential zero-bin subtraction included gives the integral
\begin{align}\label{eq:softintegral2}
I_2&=  \mu_{s}^{2\epsilon} \iota^{\epsilon} \!
  \int\!\! \frac{d^{d}q}{(2\pi)^d}
  \frac{(q^0)^2}{(q^0\!+\!i\varepsilon)^2(q^0\!-\!i\varepsilon)^2}
 \Big[  T_{s2}(q_0^2,{\bf q},{\bf k}) - T_{s2}(0,{\bf q},{\bf k}) \Big]
\nn\\
 &=  \mu_{s}^{2\epsilon} \iota^{\epsilon} \!
  \int\!\! \frac{d^{d}q}{(2\pi)^d} \frac{ \Big(\frac{{\bf k}^4}{4}+q^2{\bf k}^2+\frac{q^4}{2}\Big)}{[q^2+i\varepsilon][ (q+{\bf k})^{2}+i\varepsilon] [q_0+i\varepsilon]^2}
 = \frac{i{\bf k}^2}{16\pi^2} \bigg(-\frac{1}{\epsilon_\text{uv}}   -\ln\frac{\mu_{s}^2}{{\bf k}^2} \bigg)
 +{\cal O}(\epsilon)
 \,.
\end{align}
In the first line there is no longer a pole at $q^0=\pm i\varepsilon$ since the term in square brackets vanishes as $(q^0)^2$. Therefore we can set $(q^0)^2/(q^0-i\varepsilon)^2=1$, after which the $T_{s2}(0,{\bf q},{\bf k})/(q^0+i\varepsilon)^2$ term integrates to zero, leaving the result on the second line. (The same result is obtained if we had instead used $(q^0)^2/(q^0+i\varepsilon)^2=1$.) 

It is also reasonable to ask whether the $I_{2}$ integral could depend on the prescription adopted for the Wilson lines in the definition of $B^\mu$ in \eq{Bdefn}. An alternate definition that still leaves $B$ as an octet field was given in \eq{Bdefnalt}. Using this definition yields the same integrand $T_s(q,{\bf k})$ in \eq{Tsqk}, where we see that the eikonal propagators come in pairs $(q^0+i\varepsilon)(q^0-i\varepsilon)$ that are even under the interchange $i\varepsilon\to -i\varepsilon$.  Therefore both definitions give the same result in \eq{softcoulomb1}. 

Combining the results for $I_1$ and $I_2$ then gives the final result
\begin{align}
\label{eq:softcoulomb1}
{\rm Fig.}\ref{fig:oneloop} &=\frac{2g_{s}^4 \mu_{s}^{2\epsilon} \iota^{\epsilon} C_A}{{\bf k}^4} 
 \, (T^{A}\otimes\overline{T}^{A})\, 
 \big[ I_1 + I_2  \big] 
  \,,\nn\\
 &= \frac{i\alpha_{s}^2(\mu_s)}{ {\bf k}^2} 
  (T^{A}\otimes\overline{T}^{A})
  \bigg[ -\frac{11 C_A}{3}\Big(\frac{1}{\epsilon_\text{uv}} + \ln\frac{\mu_{s}^2}{ {\bf k}^2}\Big)-\frac{31 C_A}{9} \bigg] 
  \,,
\end{align}
which is precisely the expected result for both the $1/\epsilon_{\rm uv}$ and constant terms from the soft region at this order. 

Combining \eq{softcoulomb1} with the contribution from $n_f$ soft fermions (which using $\Xi_q$ is still the same calculation as in Ref.~\cite{Luke:1999kz}), we get the full beta function of QCD, $\beta_0 = 11 C_A/3-2 n_f/3$, which runs the Coulomb potential at one-loop,
\begin{align}
 \nu\frac{\partial}{\partial \nu}  {\cal V}_c^{(T)} &= - 2 \beta_0 \alpha_s^2(m\nu) + \ldots 
\end{align}
Note that we reproduce the correct running with the factor of $11 C_A/3$ despite the absence of ghosts in \eq{Lsintnew}.\footnote{This was also observed for the running of the Glauber potential from soft loops in Ref.~\cite{Rothstein:2016bsq}.}  This occurs because the $B$ two point function itself is gauge invariant.   Beyond one loop, the ghosts in \eq{Ls} will contribute to internal loops, and hence will appear at intermediate steps when determining the ${\cal O}(g_s^2)$ correction to \eq{BBTOP}. But once all contributions are added up the result for the $B$ two point function will again be gauge invariant. It should also be noted that  all of  the ${1}/{\epsilon}$ poles in \eqs{softintegral1}{softintegral2} are UV in origin as indicated by writing $\epsilon_{\rm uv}$.  Naively one may have thought there could be IR poles in the integral $I_2$, but these cancel out between the terms in the numerator. In general if there are $1/\epsilon_{\text{IR}}$ poles in a soft loop graph they  are converted to  $1/\epsilon_{\rm uv}$ divergences by ultrasoft 0-bin subtractions~\cite{Manohar:2006nz} (which field theoretically implements the pull-up mechanism~\cite{Hoang:2001rr}). However there are no ultrasoft zero-bin subtractions in these diagrams.  This is natural because the ultrasoft gluons decouple from the potentials at leading order. The easiest way to see this is to note that we can make a field redefinition to decouple ultrasoft gluons at leading power from the $D^0$ term in \eq{Lus}, which is analogous to the BPS field redefinition~\cite{Bauer:2001yt} in SCET,
\begin{align}  \label{eq:BPS}
 \psi_{ {\bf p}}^\prime(x) &= Y_{\rm v}(x) \psi_{ {\bf p}}(x) \,,
 \qquad
 \chi_{ {\bf p}}^\prime(x) = \bar Y_{\rm v}(x) \chi_{ {\bf p}}(x) \,,
  \\ \nonumber
 Y_{\rm v}(x) & = \exp\bigg( { -ig_u\int_{-\infty}^0\!\!\! d\lambda\ {\rm v}\cdot A^{\text{us}}(x+ \lambda {\rm v})} \bigg) \,,
\end{align}
where $Y_{\rm v}$ is an ultrasoft Wilson line. Since $Y_{\rm v}^\dagger Y_{\rm v}=1$ and $Y_{\rm v}^\dagger T^A Y_{\rm v} \otimes \bar Y_{\rm v}^\dagger \bar T^A \bar Y_{\rm v} = T^A \otimes \bar T^A$, the ${\rm v}\cdot A_{us}$ interactions decouple for both singlet and octet potentials that are independent of ultrasoft momenta.

\subsection{Simplifying the Running of the $O(v^2)$ potential}

Next we look at the running of $\mathcal{O}(v^2)$ potential in NRQCD at one loop. In vNRQCD the running of these subleading operators was performed in Refs.~\cite{Manohar:2000hj, Manohar:1999xd, Hoang:2002yy} in the non-gauge invariant basis given in (\ref{eq:Lsint}), which involved many more operators due to the need to keep track of the time-like polarizations and their associated Feynman rules. Here we will repeat the calculation of time-ordered products involving $U_{ij}^{(\sigma)} U_{ij}^{(\sigma')}$ and $W_{ij}^{(\sigma)} W_{ij}^{(\sigma')}$ with $\sigma+\sigma'=2$ in order to illustrate the advantages of our gauge invariant operator basis in \eqs{Lsintnew}{UWZ2}. Once again, the calculation of the graphs with soft quark loops do not differ from those done earlier, and are not repeated.

The contribution from the time ordered product with two $U_{ij}^{(1)}$ vertices in \fig{oneloop} is
\begin{align}
I_{11}^{U}&=\frac{i\alpha_{s}^2C_{A}}{m^2} T^{A}\otimes\overline{T}^{A}
  \bigg[\frac{15-c_{F}^2}{6}+\frac{c_{F}^2}{9}S^2
   -\frac{5( {\bf p}^2\!+\!{{\bf p}}^{\prime 2})}{{\bf k}^2}
   +\frac{5c_{F}^2}{18}T({\bf k})
   +\frac{16c_{F}}{3}\Lambda( {\bf p}^\prime, {\bf p})
    \bigg] \Big(\frac{1}{\epsilon_\text{uv}}
       +\ln\frac{\mu_s^{2}}{{\bf k}^{2}}\Big) 
  \nn\\
 &\quad + {\cal O}(\epsilon^0)
  \,,
\end{align}
where ${\bf k} ={\bf p}^\prime -{\bf p}$ and ${\cal O}(\epsilon^0)$ terms are not displayed.  The contribution from graphs with one $U_{ij}^{(2)}$ and one $U_{ij}^{(0)}$ on either vertex is
\begin{align}
I_{20}^{U}=\frac{i\alpha_{s}^2C_{A}}{m^2} T^{A}\otimes\overline{T}^{A}
 \bigg[-\frac{4}{3}\frac{({\bf p}^2\!+\!{\bf p}^{\prime 2})}{ {\bf k}^2}
  +\frac{24+5c_{F}^2}{6}
  +\frac{11c_{S}-6c_{F}}{3}\Lambda({\bf p}^\prime,{\bf p})
  \bigg] \Big(\frac{1}{\epsilon_\text{uv}}
       +\ln\frac{\mu_s^{2}}{{\bf k}^{2}}\Big)
  + {\cal O}(\epsilon^0) .
\end{align}
Since $W_{ij}^{(0)}=0$ we only have a contribution from two  $W^{(1)}$ insertions which give
\begin{equation}
I_{11}^{W}=\frac{i\alpha_{s}^2}{m^2} 
  \bigg[   \frac{14}{3}C_{1}(1\otimes\overline{1})
 -\frac{7}{6}C_d(T^{A}\otimes\overline{T}^{A})\bigg] 
  \Big(\frac{1}{\epsilon_\text{uv}}+\ln\frac{\mu_s^{2}}{{\bf k}^{2}}\Big)
   + {\cal O}(\epsilon^0) 
  \,, 
\end{equation}
where $C_1=(N_c^2-1)/(4 N_c)$ and $C_d=N_c-4/N_c$. Adding up all the contributions we find the known result \cite{Manohar:2000hj}
\begin{align}
\label{softm2}
I_{tot}=i\frac{\alpha_{s}^2}{2m^2}\Big(\frac{1}{\epsilon_\text{uv}}+\ln\frac{\mu_s^{2}}{ {\bf k}^{2}}\Big)
 &\bigg\{
 \frac{14}{3}C_{1}(1\otimes\overline{1})+(T_{A}\otimes\overline{T}_{A})
  \bigg[C_{A}\Big(\frac{39+4c_{F}^{2}}{6}
   -\frac{19}{3}\frac{{\bf p}^2+{\bf p}^{\prime 2}}{{\bf k}^2}
   +\frac{c_{F}^2}{9}S^2
    \nonumber \\
 &+\frac{5c_{F}^2}{18}T({\bf k})+ \frac{10c_{F}
  +11c_{S}}{3}\Lambda({\bf p}^\prime,{\bf p})\Big)-\frac{7}{6}C_{d}\bigg]
 \bigg\} \,,
\end{align}
which contributes a soft-loop contribution to the running of the potential coefficients ${\cal V}_i^{(0,T)}$ in \eq{V}.

\begin{figure}[t]
\centering
\includegraphics[scale=0.6]{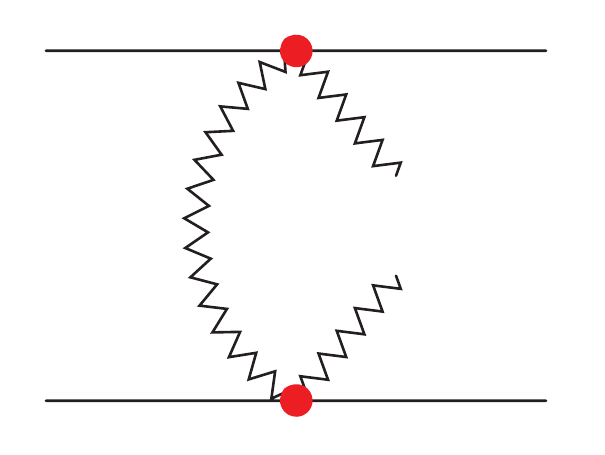}
\raisebox{-0.3cm}{
\includegraphics[scale=0.6]{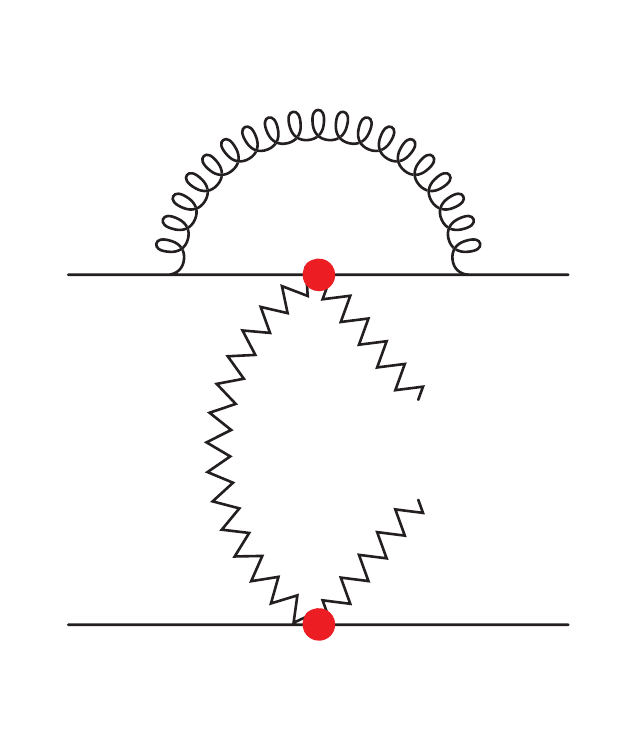}
}
\includegraphics[scale=0.6]{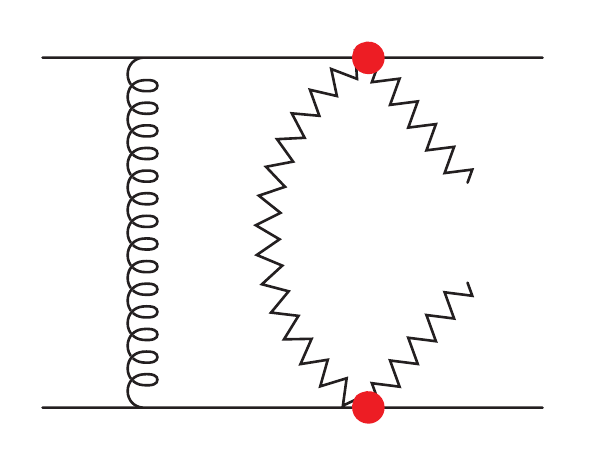} 
\includegraphics[scale=0.8]{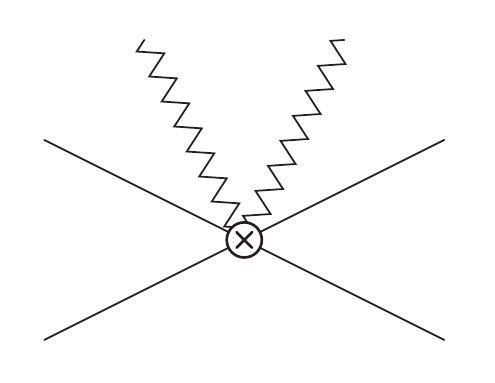} \\[-5pt]
(a) \hspace{3.2cm} (b) \hspace{3.2cm} (c) \hspace{3.2cm} (d) \phantom{x} \\
\caption{The zigzag lines denote soft gluons or quarks and curly lines are ultrasoft gluons. The time ordered product of soft fields in (a) is renormalized by ultrasoft loop graphs like those in (b) and (c), with the counterterm operator for Compton scattering off a potential shown in (d). }
\label{fig:ultrasoft}
\end{figure}

Additional contributions to running of the  spin independent part of $\mathcal{O}(v^2)$ potential come from direct ultrasoft renormalization of the potentials in \eq{V} and from ultrasoft renormalization of soft time ordered products. Since the latter involve the soft sector, we reconsider them here.  Ultrasoft modes cannot contribute in the one body sector \cite{Manohar:2000hj},
instead they renormalize  the time ordered product of two or more soft vertices with quarks and antiquarks as in \fig{ultrasoft}(a) through graphs like those in \fig{ultrasoft}(b) and \fig{ultrasoft}(c). These products appear local as far as ultrasoft gluons are concerned and it is only these products which affect observables. For a complete discussion of this subject we refer the reader to \cite{Hoang:2002yy,Pineda:2001ra}. Here we are only interested in showing that results obtained in \cite{Hoang:2002yy} follow in our manifestly soft gauge invariant formulation.
To setup the ultrasoft renormalization from graphs involving the time ordered product of two $\sigma=0$ vertices from ${\cal L}_s^{\rm int}$ in \eq{Lsintnew} we define the operators
\begin{align} \label{eq:STops}
O_{2B}^{(0,T)} &= g^{4}_{s}\mu_{s}^{4\epsilon}
  \big(\psi^{\dagger}_{\bf{p}^\prime}T^{a}\psi_{\bf{p}}\big)
  \big(\chi^{\dagger}_{\bf{-p}^\prime}\overline{T}^{b}\chi_{\bf{-p}}\big)
  \Big(B^{ic}_{-q}\: B^{jd}_{q}\Big)\, \Gamma_{B,abcd}^{(0,T)ij} 
   \,, \\
O_{2B}^{(0,1)} &= g^{4}_{s}\mu_{s}^{4\epsilon}
  \big(\psi^{\dagger}_{\bf{p}^\prime}\, \psi_{\bf{p}}\big)
  \big(\chi^{\dagger}_{\bf{-p}^\prime}\, \chi_{\bf{-p}}\big)
  \Big(B^{ic}_{-q}\: B^{jd}_{q}\Big)\, \Gamma_{B,cd}^{(0,1)ij} 
   \,,\nn \\
O_{2\Xi}^{(0,T)} &= g^{4}_{s}\mu_{s}^{4\epsilon}
  \big(\psi^{\dagger}_{\bf{p}^\prime}T^{a}\psi_{\bf{p}}\big)
  \big(\chi^{\dagger}_{\bf{-p}^\prime}\overline{T}^{b}\chi_{\bf{-p}}\big)
  \Big(\bar \Xi_{-q}\: \Gamma_{\Xi,ab}^{(0,T)}\, \Xi_{q}\Big)  
   \,, \nn \\
O_{2\Xi}^{(0,1)} &= g^{4}_{s}\mu_{s}^{4\epsilon}
  \big(\psi^{\dagger}_{\bf{p}^\prime}\, \psi_{\bf{p}}\big)
  \big(\chi^{\dagger}_{\bf{-p}^\prime}\, \chi_{\bf{-p}}\big)
  \Big(\bar\Xi_{-q}\: \Gamma_{\Xi}^{(0,1)} \,\Xi_{q}\Big)
   \,,\nn 
\end{align}
where the $\Gamma$'s are functions of the momenta, for example $\Gamma_{B,abcd}^{(0)(T)ij}=\Gamma_{B,abcd}^{(0)(T)ij}({\bf p,p'},q)$. Note that we do not have an operator in \eq{STops} with soft ghost fields, unlike Ref.~\cite{Hoang:2002yy}.
These structures are determined by the form of the time-ordered products in \fig{ultrasoft}(a), so for example the structure in the first operator is:
\begin{align}
\Gamma_{B,abcd}^{(0,T)ij}
 =&-\frac{f_{aec}f_{be'd}}{2}\Big[ 
 U^{(0)ii'}_{q,\bf{p},\bf{p^\prime}}U^{(0)jj'}_{-q,\bf{-p},\bf{-p^\prime}}
(-i)\int\!\! d^4{q^\prime}\, \langle0| B^{i'e}_{q+{\bf k}}B^{j'e'}_{q^\prime} |0\rangle\Big] 
\nonumber \\ 
 &-\frac{f_{aed}f_{be'c}}{2}\Big[
 U^{(0)ii'}_{-q,\bf{p},\bf{p^\prime}}U^{(0)jj'}_{q,\bf{-p},\bf{-p^\prime}}
 (-i)\int\!\! d^4q^\prime\, \langle0| B^{i'e}_{q-{\bf k}}B^{j'e'}_ {q^\prime}|0\rangle\Big] \,,
\end{align}
where as usual ${\bf k=p'-p}$ and the time ordered product of $B$ fields is defined in \eq{BBTOP}. 
\begin{figure}
	\centering
	\includegraphics[scale=0.8]{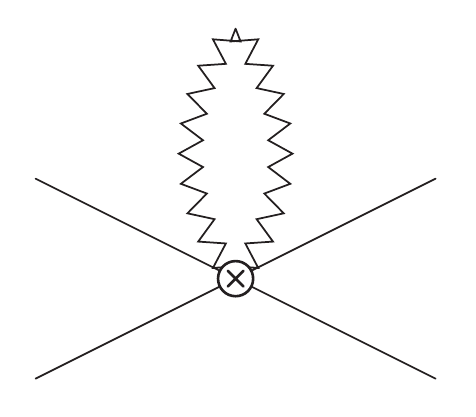}
	\caption{This diagram contributes to the running of spin-independent part of $O(v^2)$ vNRQCD potential.}
	\label{fig:ultrasoftrunning}
\end{figure}

The Wilson coefficients of the operators in \eq{STops} vanish at tree level, since
the analogous contributions are reproduced directly by a soft time ordered product in the EFT,  but if we consider an ultrasoft gluon exchange between the fermions (examples are given in Fig.\ref{fig:ultrasoft}(b,c)) then we can generate non-zero Wilson coefficients for these operators by mixing.  Here the ultrasoft gluons can attach to any of the heavy fermion lines.  These operators do not affect the running of the leading order potential since non-canceling ultrasoft gluon graphs bring at least an extra factor of $v^2$ (for example via the dipole vertex ${\bf p}\cdot {\bf A}_{u}$, see \eq{Lus}).  The operators which get renormalized by  ultrasoft loop graphs, like those shown in  Fig.{\ref{fig:ultrasoft}(b,c)}, are 
\begin{align}
\label{eq:ultrasoft}
O^{(2,1)}_{2a}&=\frac{{\bf k}^2}{m^2} 
   \big(O^{(0,1)}_{2B}+O^{(0,1)}_{2\Xi} \big) 
   \,,\nonumber \\
O^{(2,T)}_{2b}&=\frac{{\bf k}^2}{m^2}
   \big(O^{(0,T)}_{2B}+O^{(0,T)}_{2\Xi} \big) 
   \,,\nonumber \\
O^{(2,T)}_{2c}&=\frac{({\bf p}^2+{\bf p}^{\prime 2})}{m^2} 
   \big( O^{(0,T)}_{2B}+O^{(0,T)}_{2\Xi} \big)
    .
\end{align} 
Once again the results for the diagrams involving soft quark propagators are identical to earlier results at the order of our computation. 
By expanding the $B_\mu$ field in the $O_{2B}$ operators for Fig.(\ref{fig:ultrasoft})(d) to leading order in $g$, we can show that our results match directly with results in \cite{Hoang:2002yy}. Since ultrasoft modes do not couple to soft modes, an ultrasoft loop will generate the same Wilson coefficient for the operators with $B^\mu$ fields as it would if we used $A^\mu$ fields instead of $B^\mu$ fields. Thus we may extract the renormalized Wilson coefficients for the operators in \eq{ultrasoft}, which after renormalization group evolution in vNRQCD yields~\cite{Hoang:2002yy}:
\begin{align}
\label{Wilsoncoeff}
C_{O^{2,T}_{2c}}(\nu)&=\frac{-4C_{A}}{3\beta_{0}}\ln(w) 
  \,,  \\
C_{O^{2,T}_{2b}}(\nu)&=\frac{3C_{A}-C_{d}-4C_{F}}{3\beta_{0}}\ln(w) 
  \,, \nonumber \\
C_{O^{2,1}_{2a}}(\nu)&=\frac{4C_{1}}{3\beta_{0}}\ln(w)
  \,, \nonumber
\end{align}
where $w={\alpha_{s}(m\nu^2)}/{\alpha_{s}(m\nu)}$.  These coefficients contribute to the running for $O(v^2)$ spin-independent vNRQCD potentials via the diagram shown in Fig.(\ref{fig:ultrasoftrunning}), see~\cite{Hoang:2002yy} and \cite{Pineda:2001ra}.

\section{Higher Order Loop Corrections to the QCD Potential}
\label{sec:Vctwoloop}

In the full theory the static potential $V(R)$ is often defined by the time like Wilson loop
\beq \label{eq:wilson}
  \langle 0 \mid P e^{i\oint dt A_0(t,\vec x)} \mid 0 \rangle = e^{i V(R) T} .
\eeq
In vNRQCD the leading power Coulomb-like potential is obtained from a combination of the ${\cal V}_c^{(1,T)}$ potential matching coefficients in \eq{V} and time ordered products involving soft fields.  The soft time ordered products also induce the running of the strong coupling in the potential.  In the Wilson loop calculation of $V(R)$ using \eq{wilson}, at three loops one encounters an infra-red divergence known as the ADM singularity~\cite{Appelquist:1977es}. However, this divergence is not part of the potential \cite{Brambilla:1999qa}, as it arises from the ultrasoft region of
integration and thus must be subtracted. In the vNRQCD formalism this divergence is removed by an ultrasoft zero-bin subtraction to the soft loop computation~\cite{Manohar:2006nz}.  The two loop static potential has been calculated in Refs.~\cite{Peter:1996ig, Schroder:1998vy}, and the three loop soft potential has been calculated  in Refs.~\cite{Smirnov:2008fk,Smirnov:2009uq,Anzai:2009tm, Smirnov:2010vn}. 

An interesting aspect of the two loop static potential, pointed out in \cite{Schroder:1998vy}, is that the $i\varepsilon$'s in the propagators of soft heavy quarks must be accounted for to obtain the correct result. In our one loop discussion we remarked that the  sign of these $i\varepsilon$'s only contributes a term where the energy of a soft gluon goes to zero, i.e a potential exchange that is removed by the zero-bin subtractions. At two loops, the calculation in \cite{Schroder:1998vy} indicates that this is no longer the case, and we will explore how this relates to our formalism with $B$ operators in this section. 

In particular, it should be noted that we do not have complete freedom to pick the $i\varepsilon$'s that appear in Feynman rules associated with $B^\mu$, because the allowed structure of Wilson lines is constrained by requiring that $B^\mu =B^{\mu a}T^a$ is purely an octet field, as in \eq{Boctet}, which in turn requires  the use of the identity $S_{\rm v}^\dagger S_{\rm v}=1$ and, for example, would not be the case if we tried to use $S^{\dagger}_{\rm  v}(x,-\infty)\, iD_s^\mu (x)\,  S_{\rm  v}(x,+\infty)$ to define $B^\mu$.  This is consistent with the fact that not all $i\varepsilon$'s from the soft Wilson lines in $B^\mu$ lead to iterated potentials. As we will see below, for some of these terms there is no zero-bin subtraction and therefore the soft loop integrals have non-trivial contributions where these $i\varepsilon$'s are important.

\begin{figure}[t]
	\centering
	\includegraphics[scale=0.5]{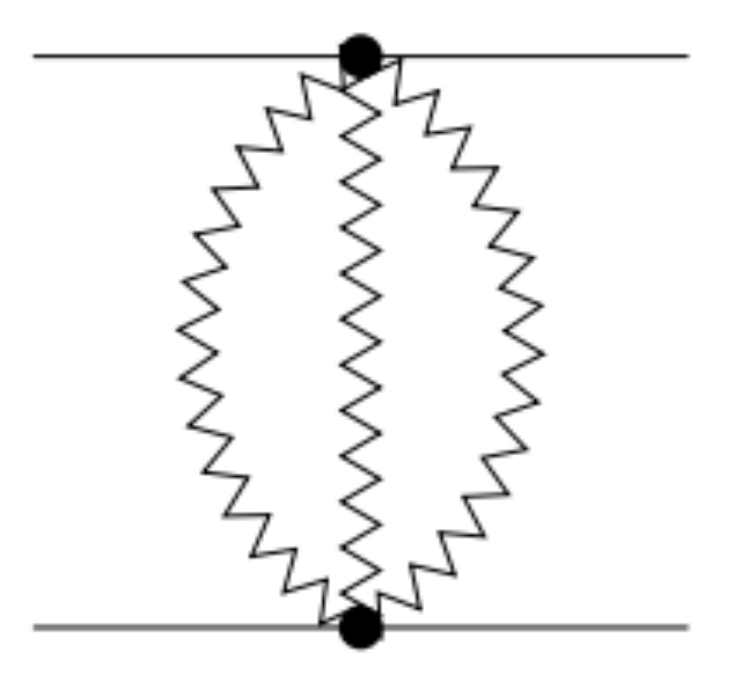}
	\caption{Two loop EFT diagram for running of the leading order potential that arises when using the gauge potential as the canonical variable and this absent when quantizing using $B$.}
	\label{twoloop1}
\end{figure}

Working with the $B^\mu$ fields in the  EFT introduces some apparent technical complications that we will now show are straightforward to handle. In particular, when using \eq{wilson} any integral that is ill defined due a pinch in an energy contour integral, need not be calculated since non-abelian exponentiation~\cite{Gatheral:1983cz,Frenkel:1984pz} implies that they are iterations of lower order potentials terms from the matrix element in \eq{wilson} that are proportional to $T^{k\ge 2}$, where $T$ is defined in \eq{wilson} . In contrast, in vNRQCD the pinch singularities in soft loop calculations are removed by potential zero-bin subtractions~\cite{Manohar:2006nz}. Also, the contributions which need not be calculated in the Wilson loop calculation, due to non-abelian exponentiation, do not even appear from the vNRQCD Feynman diagrams generating purely soft loops. However, in our formalism using $B$'s the non-trivial soft loop contributions can themselves come with pinch singularities, unlike results obtained from static potential calculations starting with \eq{wilson}.  Thus in the EFT we must deal with pinched integrals to calculate the potential, and the corresponding zero-bin subtractions.

As an example of how to properly perform the zero-bin subtractions, consider the two loop integral which arises from the  diagram in Fig.~\ref{twoloop1}. We will neglect the  terms in the integrand that do not have the maximal number of time-like $k^0$ propagator factors since this allows us to focus on the most complicated term, which has overlapping subtractions.  After expanding out the $B$ fields in terms of soft gluon fields to higher order using \eq{Bexpn}, and performing the contractions, the relevant integral is  
\begin{align}
\label{eq:twoloopint} 
I=i\frac{g_s^6\mu_s^{6\epsilon}\iota^{3\epsilon}}{\bf k^4}
  \frac{C_A^2}{6}(T^e\otimes  \bar T^e) \int 
  & \frac{[d^dk_1]}{(k_1^2+i\varepsilon)}
    \frac{[d^dk_2]}{(k_2^2+i\varepsilon)}
    \frac{({\bf k}^2 -{\bf k}_1^2-{\bf k}_2^2-{\bf k}_3^2)^2}{(k_3^2+i\varepsilon)}
    \bigg[ \frac{1}{(-k_1^0)^2_A(-k_3^0)^2_A}
  \nn \\ 
  &  +\frac{1}{(-k_2^0)^2_A(-k_3^0)^2_A} 
     +\frac{1}{(-k_1^0)_A(-k_2^0)_A(-k_3^0)_A^2}\bigg],
\end{align}
where $[d^dk_i]=d^dk_i/(2\pi)^d$, $k=(0, {\bf k})$ is the net momentum exchange between heavy quarks and 
\begin{align}
 k_3\equiv -k_1-k_2+k \,.
\end{align}
We have also defined
\begin{align} \label{eq:k0A}
\frac{1}{(-k_i^0)_A}\equiv \frac{k_i^0}{(k_i^0+i \varepsilon)(-k_i^0+i \varepsilon)}= -\frac{1}{2}\left(\frac{1}{(k_i^0+i \varepsilon)}-\frac{1}{(-k_i^0+i \varepsilon)}\right) \,,
\end{align}
which is antisymmetric under $k_i^0\to -k_i^0$. Once again we find that this same result is obtained whether we start with \eq{Bdefn} or \eq{Bdefnalt} as the definition for $B$. 
For the terms in the numerator involving the loop momenta, ${\bf k}_i^2$, we can write ${\bf k}_i^2 = (k_i^0)^2 - k_i^2$ and cancel either eikonal or relativistic propagators. For simplicity we will only examine the term with the external momentum ${\bf k}^4$ in the numerator where no propagators are canceled. 

Let us simplify this term in the integrand by putting it into a form which makes all the pinch singularities manifest. We will use  the notation in \cite{Schroder:1998vy},  introducing
\begin{align}
S_{i}=\frac{1}{k^{0}_{i}+i\varepsilon} \,,\quad
S_{\overline{i}}=\frac{1}{-k^{0}_{i}+i\varepsilon} \,,\quad
~~S_{ij\cdots}=S_iS_j\cdots
 \,,
\end{align}
and defining
\begin{align} \label{eq:k5}
k_5 \equiv k_1 - k_2 \,.
\end{align} 

Making the change of variables
\beq \label{eq:cov}
 k_{1}\rightarrow k_{2},\qquad k_{2}\rightarrow k_1-k_2
\eeq 
such that  $k_{3}\rightarrow k-k_{1}$, and using \eq{k0A} we can simplify the first term in the integrand in Eq.~(\ref{eq:twoloopint}) as follows: 
\begin{align} \label{eq:doublebox} 
\frac{1}{(-k_1^0)^2_A(-k_3^0)^2_A}
 &= \frac{1}{16}\big( S_1 - S_{\bar 1} \big)^2 \big( S_2 - S_{\bar 2} \big)^2
 \nn \\
 & \!\!\!\!\!\!\!
 =\frac{1}{16}\Big(S_{1122}+S_{\overline{1}\overline{1}22}+S_{11\overline{2}\overline{2}}  + S_{\overline{1}\overline{1}\overline{2}\overline{2}}
  -2S_{1\overline{1}22}-2S_{112\overline{2}}-2S_{1\overline{1}\overline{2}\overline{2}}-2S_{\overline{1}\overline{1}2\overline{2}}+4S_{1\overline{1}2\overline{2}}
			\Big) 
\nn\\
 & \!\!\!\!\!\!\!
 =\frac{1}{8}\left(S_{1122}+S_{11\overline{2}\overline{2}}-4S_{112\overline{2}}+2S_{1\overline{1}2\overline{2}}\right)
 \,.
\end{align}
The last equality follows from the symmetry of the integral under $v\rightarrow{-v}$ and under $\{k_{1}\to k-k_2, k_2\to k-k_1\}$. The last term in Eq.~(\ref{eq:doublebox}) has a double pinch singularity, and the second to last term has a single pinch singularity, whereas the first two terms have no pinch singularities. All four terms will have zero-bin subtractions, but the zero-bins of the first
two terms will vanish upon contour integration. The $i\varepsilon$ factors in the $S_{1122}$ and $S_{11\overline{2}\overline{2}}$ are nevertheless important for evaluating these integrals. In contrast, the last two terms in have \eq{doublebox} have subtractions that must be included, which we can check by examining their power counting in the potential limit.

Consider the scaling of the term $S_{1\overline{1}2\overline{2}}$ in Eq.~(\ref{eq:doublebox}) which has a double pinch singularity, and call its full integral $\tilde J$ prior to including any subtractions. Here $\tilde J\sim 1/v^2$ is the scaling of this integral when both loop momenta are soft.
The relevant zero-bin limits to consider and resulting scaling for the corresponding zero-bin loop integrals $J_0^{(X)}$ are
\begin{align} \label{eq:Jscaling}
 & k_1^0\sim v^2,\ k_2^0\sim v: & J_0^{(k_1)} \sim \frac{1}{v^3}  \,,\\
 & k_1^0\sim v,\ k_2^0\sim v^2: & J_0^{(k_2)} \sim \frac{1}{v^3} \,,\nn\\
 & k_1^0\sim v^2,\ k_2^0\sim v^2: & J_0^{(k_1k_2)} \sim \frac{1}{v^4} \,.
 \nn
\end{align}
From this we conclude that all three of these zero-bin subtractions must be included, since $J_0^{(k_1)},J_0^{(k_2)}\sim 1/v^3$ scale in the same way as a two loop graph with a potential loop generated by the iteration of one soft loop and one tree level potential, and $J_0^{(k_1k_2)}\sim 1/v^4$ scales in the same way as a graph with two potential loops generated by the iteration of three tree level potentials. For example, to construct the first integral we take $k_1^0\sim v^2$, $k_2^0\sim v$ and hence expand the integrand of $\tilde J$ with $k_1^0\ll k_2^0$, ${\bf k}_1$, ${\bf k}_2$. This gives the zero-bin integral
\begin{align}
\label{eq:zb}
J_{0}^{(k_1)} &= \!\int\!\! \frac{[dk_1^0 d^{d-1}{\bf k}_1][dk_2^0 d^{d-1}{\bf k}_2]}
 {(k_2^2+i\varepsilon)(-{\bf k}_3^2)[(k_2^0)^2 -({\bf k}_1 -{\bf k}_2)^2+i\varepsilon]} 
 \frac{1}{(k^0_1+i\varepsilon)} \frac{1}{(-k^0_1+i\varepsilon)}
\frac{1}{(k^0_2+i\varepsilon)} \frac{1}{(-k^0_2+i\varepsilon)}
 \nn\\
 &\sim \frac{1}{v^3}  \,.
\end{align}
Subtracting Eq.~(\ref{eq:zb}) from last term of Eq.~(\ref{eq:doublebox}) makes the remaining integral well defined with respect to the $k_{1}^{0}$ integral. In $J_0^{(k_1)}$ there is still a contribution from the region where the second loop momentum scales into the potential region, $k_2^0\sim v^2$, which must be subtracted by a term $J_0^{(k_1)(k_2)}$ that is constructed in an analogous manner but starting with $J_0^{(k_1)}$. The result for $J_0^{(k_2)}$ is analogous to \eq{zb}, and it has a contribution from $k_1^0\sim v^2$ that requires a subtraction $J_0^{(k_2)(k_1)}$. This discussion follows the standard zero-bin approach for multiloop integrals with overlapping subtractions. These terms give the same integral as the third $k_1^0\sim k_2^0\sim v^2$ subtraction listed in \eq{Jscaling}. This integral is 
\begin{align}
 J_0^{(k_1k_2)} &=J_0^{(k_1)(k_2)}=J_0^{(k_2)(k_1)}
  \\
   &=  \!\int\!\! \frac{[dk_1^0 d^{d-1}{\bf k}_1][dk_2^0 d^{d-1}{\bf k}_2]}
   {(-{\bf k}_2^2)({\bf k}_3^2)({\bf k}_1 -{\bf k}_2)^2} 
   \frac{1}{(k^0_1+i\varepsilon)} \frac{1}{(-k^0_1+i\varepsilon)}
   \frac{1}{(k^0_2+i\varepsilon)} \frac{1}{(-k^0_2+i\varepsilon)}
    \,. \nn
\end{align}
Thus the  net effect of including $J_0^{(k_1)(k_2)}$ and $J_0^{(k_2)(k_1)}$ is simply to flip the sign in front of the $J_0^{(k_1k_2)}$ subtraction. For the total integral combining these terms then gives
\begin{align}
  J = \tilde J - J_0^{(k_1)} -J_0^{(k_2)} + J_0^{(k_1k_2)}
\end{align}
Altogether this leads to a combined result that has well defined integrals in both $k_1^0$ and $k_2^0$.  Although here we are discussing graphs that contribute to the $C_A^2 (T^e\otimes \bar T^e)$ color structure, the nature  of the zero-bin subtractions for the $S_{1\overline{1}2\overline{2}}$ contribution is actually identical to the two loop double box graph in the HQET based loop computation of the potential. 

The $S_{112\overline{2}}$ term in Eq.~(\ref{eq:doublebox}) is even simpler to handle since it only has a single non-zero zero-bin subtraction for $k_2^0\sim v^2$. The analysis for $S_{112\overline{2}}$ is identical to the two loop graph that is a cross box with an extra gluon rung added.  Thus all pinch singularities in Eq.~(\ref{eq:doublebox}) are converted to well defined integrals once the zero-bin subtraction terms are included. 

Note that the first two terms inside the square bracket in Eq.~(\ref{eq:twoloopint}) are equal by the symmetry of the original integral under $k_{1}\leftrightarrow k_{2}$, so our discussion so far also suffices to cover the second term in this expression. 

Finally we consider the last term in Eq.~(\ref{eq:twoloopint}) as this demonstrates a case where there is no analog of the required potential zero-bin subtractions in the Wilson loop based calculation. First we make the change of variables in \eq{cov}, and use the $v\to -v$ symmetry to write
\begin{align} \label{eq:doublebox3}
\frac{1}{(-k_1^0)_A(-k_2^0)_A(-k_3^0)^2_A}&= \frac{1}{16}(S_{2}-S_{\overline{2}})(S_{5}-S_{\overline{5}})(S_{\overline{1}}-S_{1})^2 \nn \\
&= \frac{1}{16}(S_{11}+S_{\overline{1}\overline{1}}-2S_{1\overline{1}})(S_{25}+S_{\overline{2}\overline{5}}-S_{2\overline{5}}-S_{\overline{2}5})
\nn \\
&=\frac{1}{8}(S_{11}+S_{\overline{1}\overline{1}}-2S_{1\overline{1}})(S_{25}-S_{2\overline{5}}) 
 \nn\\
&=\frac{1}{8}\big(
  S_{1125}- S_{112\overline{5}} -S_{11\overline{2}5}
  -2S_{1\overline{1}25}
  +2 S_{1\overline{1}2\overline{5}}
  +S_{\overline{1}\overline{1}25}
  \big)
\,,
\end{align}
where $k_5$ was defined in \eq{k5}. In the final expression here the first three terms do not have pinch singularities, and can be directly computed using standard two-loop techniques. In general the $i\varepsilon$ factors in the eikonal propagators are needed to obtain correct results for these terms. In contrast, the last three terms in \eq{doublebox3} have pinch singularities. This is immediately apparent for the fourth and fifth terms which involve a $S_{1\overline{1}}$. For the sixth term the pinch singularity is revealed after performing one of the energy integrals, for example integrating in $k_2^0$ yields an answer with a pinch singularity in $k_1^0$.  Since the nature of the zero-bin subtractions for the last three terms in \eq{doublebox3} is quite similar, we will choose as an example to go through the analysis for $S_{\overline{1}\overline{1}25}$ with only the ${\bf k}^4$ term kept in the numerator and again pulled outside the integrand. We call the corresponding full integral $\tilde L$ prior to including any subtractions. 

Once again we first enumerate the necessary zero-bin subtractions by considering the scaling of the $\tilde L$ integrand and measure in potential limits:
\begin{align} \label{eq:Lscaling}
 & k_1^0\sim v^2,\ k_2^0\sim v: & L_0^{(k_1)} &\sim \frac{1}{v^3}  \,,\\
 & k_1^0\sim v,\ k_2^0\sim v^2: & L_0^{(k_2)} &\sim \frac{1}{v^2} \text{ (no zero-bin)} \,,\nn\\
 & k_1^0\sim v,\ k_5^0\sim v^2: & L_0^{(k_5)} &\sim \frac{1}{v^2} \text{ (no zero-bin)} \,,\nn\\
 & k_1^0\sim v^2,\ k_2^0\sim v^2: & L_0^{(k_1k_2)} &\sim \frac{1}{v^4} \,.
 \nn
\end{align}
Note that here because of the presence of a $S_5$ term that we also considered a $k_5^0\sim v^2$ potential limit. The non-zero zero-bin integrals are
\begin{align} \label{eq:zbL}
L_{0}^{(k_1)} 
 &= \!\int\!\! \frac{[dk_1^0 d^{d-1}{\bf k}_1][dk_2^0 d^{d-1}{\bf k}_2]}
 {(k_2^2+i\varepsilon)(-{\bf k}_3^2)[(k_2^0)^2 -({\bf k}_1 -{\bf k}_2)^2+i\varepsilon]} 
 \frac{1}{(-k^0_1+i\varepsilon)^2} 
\frac{1}{(k^0_2+i\varepsilon)} \frac{1}{(-k^0_2+i\varepsilon)}
 \,, \nn \\
L_{0}^{(k_1k_2)} 
 &= \!\int\!\! \frac{[dk_1^0 d^{d-1}{\bf k}_1][dk_2^0 d^{d-1}{\bf k}_2]}
   {(-{\bf k}_2^2)({\bf k}_3^2)({\bf k}_1 -{\bf k}_2)^2} 
 \frac{1}{(-k^0_1+i\varepsilon)^2} 
\frac{1}{(k^0_2+i\varepsilon)} \frac{1}{(k_1^0-k^0_2+i\varepsilon)} \,.
\end{align}
Again there is an additional subtraction made on $L_0^{(k_1)}$ to ensure its $k_2^\mu$ momentum remains soft (or equivalently, that this integral does not double count the result from $L_{0}^{(k_1k_2)}$). This term is given by
\begin{align}
 L_{0}^{(k_1)(k_2)} 
 &= \!\int\! \frac{[dk_1^0 d^{d-1}{\bf k}_1][dk_2^0 d^{d-1}{\bf k}_2]}
   {(-{\bf k}_2^2)({\bf k}_3^2)({\bf k}_1 -{\bf k}_2)^2} 
 \frac{1}{(-k^0_1+i\varepsilon)^2} 
\frac{1}{(k^0_2+i\varepsilon)} \frac{1}{(-k^0_2+i\varepsilon)}
 \,.
\end{align}
Note that in this case $L_0^{(k_1)(k_2)} \ne L_0^{(k_1k_2)}$. The final purely soft result is then obtained by 
\begin{align}
  L = \tilde L - L_0^{(k_1)} + L_0^{(k_1)(k_2)} - L_0^{(k_1k_2)} \,.
\end{align}
To evaluate this result we first note that after combining integrands the combination $-L_0^{(k_1)} + L_0^{(k_1)(k_2)}$ no longer has a pinch singularity in $k_2^0$, and has a $k_1^0$ integral that yields zero. This leaves $L=\tilde L - L_0^{(k_1k_2)}$. Putting these two integrands over a common denominator gives terms in the numerator involving energies as $\{ (k_1^0)^2, (k_2^0)^2, k_1^0 k_2^0 \}$ or with higher powers. These all yield well defined integrals. Thus the zero-bin again removes the pinch singularities, yielding a well defined result for $L$. The analysis for the $S_{1\overline{1}2\overline{5}}$ and $S_{1\overline{1}2\overline{5}}$ terms in \eq{doublebox3} is similar, with the subtractions again yielding well defined final results.

In the above analysis, only the pinched poles are removed by zero-bin subtractions, while the other $\pm i\varepsilon$ terms in the eikonal propagators are important for obtaining the final results for integrals.   These poles can lead to additional $\pi^2$ terms, which are related to the contribution discussed by \cite{Schroder:1998vy}.  We thus find that  the precise path for the Wilson lines used to define $B$ is again not relevant at two-loops (and presumably to all orders). However, this does not imply that the signs of the $i\varepsilon$ obtained from expanding the soft Wilson lines are not needed to obtain the correct results. Indeed, it is important that $B$ is defined as an octet field. With this constraint satisfied, changing the path of the Wilson lines flips the signs of all eikonal $i\varepsilon$'s in a correlated manner, and these correlations matter. 
	
Finally we note that working in the effective theory can illuminate higher order corrections to the potential. First of all, there are fewer diagrams since there is not a term with a single soft gluon coupling to the potential quarks. The color structure is also clarified since in the EFT only terms which actually contribute to the potential arise in matching. For instance, at two loops the soft contribution of all the integrals proportional to $C_F^3$ as well as those proportional to $C_F^2C_A$ must sum to zero if we use the Wilson loop definition of the potential (or using the HQET implementation of the soft sector), whereas in the treatment advocated here no such color structures in soft integrals arise.

\section{Conclusions and Outlook}

We formulated the interaction terms between potential heavy quarks and soft fields in vNRQCD in a way which is manifestly soft and ultrasoft gauge invariant. This improves upon the construction of Ref.~\cite{Luke:1999kz} as it greatly reduces the size of the operator basis for this Lagrangian, and leads to simpler calculations for soft loop diagrams at both leading and subleading power.  Most importantly perhaps this formulation in terms of the soft gauge invariant building blocks $B$ and $\Xi$ is likely to lead to simplifications for higher order matching and anomalous dimension calculations in the soft sector.

Our formalism also helps to elucidate the role of the $i\varepsilon$ prescription in soft heavy quark propagators.  These appear from soft Wilson lines in the formalism used here, and give additional pinch singularities that are then removed by zero-bin subtractions from the potential region of momentum space.  Interestingly, the zero-bin subtractions in this analysis do not have a 1-to-1 correspondence with EFT diagrams with potential loop momenta, unlike the typical situation that is encountered for zero-bin subtractions. This is easily seen  since the zero-bin subtractions have  different color structures than the diagrams obtained by iterating potentials in the examples treated here. Another interesting point arises at two-loops, where the signs of the $i\varepsilon$'s obtained from the Wilson lines inside the soft building block field $B$ do matter for obtaining the final result for the soft loop integral, but the path chosen for these Wilson lines does not matter as long as it is chosen such that $B$ is a purely octet field. This constraint rules out some possible combinations of Wilson line paths that would have still led to a gauge invariant defintion of $B$.  Given the analogy between the use of our $B$ for vNRQCD, and the use of the soft and collinear building block fields for Glauber operators in SCET~\cite{Rothstein:2016bsq}, it seems quite likely that a similar conclusion will apply there as well.

An interesting possibility raised by the formalism developed here is to consider quantizing the soft $B^{\mu}$ and $\Xi$ fields themselves rather than the soft gauge field and soft quark field. An analogous method of quantization fields that involve collinear Wilson lines was considered for SCET in Ref.~\cite{Bauer:2008qu}, but has not yet found wide use there.\footnote{For the same reasons discussed here, it may however prove to be useful for calculations involving Glauber exchange, since these operators are formulated directly in terms of the building block fields~\cite{Rothstein:2016bsq}.} If we consider this approach in vNRQCD, then it leads to even simpler Feynman diagrams since \eq{Lsintnew} now yields four point vertex Feynman rules, but no higher point vertices.  With soft fields quantized in this manner, the two point function for $B$ is given by \eq{BBTOP} dropping terms at and beyond ${\cal O}(g^2)$ since the $B$ fields are now fundamental. Furthermore, the three and four point interactions for $B$s are identical to those for  the triple and quadruple gluon vertices in QCD~\cite{Bauer:2008qu}.  We find that considering this approach in vNRQCD is like working in a ${\rm v}\cdot A_s=0$ gauge for the soft gluon fields, utilizing zero-bin subtractions to handle the extra singularities in the gauge propagator that otherwise make this gauge complicated (see eg.~\cite{Leibbrandt:1987qv}), while systematically avoiding power counting violating corrections that are present for HQET in this gauge~\cite{Manohar:2000dt}.\footnote{Effectively, our approach implies that we do not use this gauge for calculations that would involve potential gluons since they were integrated out, but only use it for the soft gluons.} It would therefore be interesting to further explore the direct use of a quantized $B$ and $\Xi$ for carrying out soft NRQCD calculations.

Another interesting idea for further exploration is using the field redefinition in \eq{BPS} to obtain a simpler set of ultrasoft couplings to potential heavy quarks. Just like in SCET this will lead to the appearance of ultrasoft gauge invariant gluon building block fields through the identity $Y_{\rm v}^\dagger i D_{us}^\mu Y_{\rm v} = i\partial_{us}^\mu - g {B}_{us}^\mu$, where ${\rm v}\cdot {B}_{us}=0$. Here ${B}_{us}$ is defined in the same manner as \eq{Bdefn} but with the $S_{\rm v}$ Wilson lines replaced by the $Y_{\rm v}$ Wilson lines.

\begin{acknowledgments}
  We thank Andre Hoang and Aneesh Manohar for helpful comments on the draft. This work was supported in part by the U.S. Department of Energy (DOE) and the Office of Nuclear Physics under DE-SC0011090, and by DOE grants DE-FG02-04ER41338 and FG02-06ER41449.  I.S. was also supported by the Simons Foundation through the Investigator grant 327942. 
\end{acknowledgments}

\bibliography{nrqcd}{}
\bibliographystyle{jhep}

\end{document}